%% file: wRebeca.tex
\documentclass{fac}

\ifprodtf \else \usepackage{latexsym}\fi

\newcommand\black{\ensuremath{\blacktriangleright}}
\newcommand\white{\ensuremath{\vartriangleright}}

\newif\ifamsfontsloaded
\ifprodtf
  \newcommand\whbl{\white\kern-.1em--\kern-.1em\black}
  \newcommand\blwh{\black\kern-.1em--\kern-.1em\white}
  \newcommand\blbl{\black\kern-.1em--\kern-.1em\black}
  \newcommand\whwh{\white\kern-.1em--\kern-.1em\white}
  \amsfontsloadedtrue
\else
  \checkfont{msam10}
  \iffontfound
    \IfFileExists{amssymb.sty}
      {\usepackage{amssymb}\amsfontsloadedtrue
       \newcommand\whbl{\white\kern-.125em--\kern-.125em\black}%
       \newcommand\blwh{\black\kern-.125em--\kern-.125em\white}%
       \newcommand\blbl{\black\kern-.125em--\kern-.125em\black}%
       \newcommand\whwh{\white\kern-.125em--\kern-.125em\white}}
      {}
  \fi
\fi

\newcommand{\overto}[1]{\stackrel{#1}{%
		\overrightarrow{\smash{\,{\phantom{#1}}\,}}}}

\newcommand{\pverto}[1]{\stackrel{#1}{%
                   \overrightarrow{\smash{\,{\phantom{#1}}\,}}'}}
\newcommand{\Rule}[2]{                                  % operational rule
    \frac{\raisebox{.7ex}{\normalsize{$#1$}}}
    {\raisebox{-1.0ex}{\normalsize{$#2$}}}}

\newcommand{\FAxiom}[1]{                                  % operational rule
    {\normalsize {#1}}
    }

\newcommand{\push}[1]{\{\!\{#1\}\!\}}
\newcommand{\A}[1]{\mathbf{A}^{\,#1}}
\newcommand{\U}[2]{{\,}_{#1}\mathbf{U} {\!}_{#2}\,}
\newcommand{\W}[2]{{\,}_{#1}\mathbf{W} {\!}_{#2}\,}
\newcommand{\pconn}{\dashrightarrow}

\newtheorem{theorem}{Theorem}[section]
\newtheorem{definition}{Definition}[section]

\title[Modeling and Efficient Verification of Wireless Ad hoc Networks]  {Modeling and Efficient Verification of Wireless Ad hoc Networks}

\author[Behnaz Yousefi, Fatemeh Ghassemi, Ramtin Khosravi]   {Behnaz Yousefi, Fatemeh Ghassemi and  Ramtin Khosravi\\
     School of Electrical and Computer Engineering, University of Tehran, Iran \\
     \{b.yousefi,
     fghassemi,
     r.khosravi\}@ut.ac.ir}

\correspond{Christiane Notarmarco, Springer-Verlag London
            Limited, Sweetapple House, Catteshall Road,
            Godalming, Surrey GU7 3DJ, UK.
            e-mail: chris@svl.co.uk}

\pubyear{2000}
\pagerange{\pageref{firstpage}--\pageref{lastpage}}

\usepackage{amssymb}
\usepackage{graphicx}
\usepackage{caption}
\usepackage{subcaption}
\usepackage{url}
\usepackage[usenames,dvipsnames]{color}
\usepackage[usenamesdvipsnames]{xcolor}
\usepackage{mathtools}
\usepackage{perpage}
\usepackage{hyperref}
\usepackage{booktabs}
\usepackage{multicol}
\usepackage{float}
\usepackage{listings}
\usepackage{alltt}
\usepackage[figure,noline]{algorithm2e}
\usepackage{tikz}
\usepackage{bussproofs}

\lstdefinelanguage{rebeca}{
  morekeywords={reactiveclass, knownrebecs, statevars, main, msgsrv, constraint,con, main, define, LTL, CTL, boolean, int, shortint, byte, if, else, while, for, wait, msg, reset, set, self, false, true, now, after, delay, deadline, initial},
  otherkeywords={=>,<-,<\%,<:,>:,\#,@},
  sensitive=true,
  morecomment=[l]{//},
  morecomment=[n]{/*}{*/},
  morestring=[b]",
  morestring=[b]',
  morestring=[b]"""
}
\begin{document}
\label{firstpage}

\makecorrespond

\maketitle

\input{abstract}
\input{Introduction}
\input{Preliminaries}
\input{BroadcastingRebecs}

\input{ApplyingCounterAbstraction2}

\input{CaseStudies}
\input{Evaluation}
\input{RelatedWork}

\input{Conclusion}
\bibliographystyle{alpha}
\bibliography{ref}
\label{lastpage}

\end{document}

%% file: abstract.tex
\begin{abstract}
Wireless ad hoc networks, in particular mobile ad hoc networks
(MANETs), are growing very fast as they make communication easier
and more available. However, their protocols tend to be difficult to
design due to topology dependent behavior of wireless communication,
and their distributed and adaptive operations to topology dynamism.
Therefore, it is desirable to have them modeled and verified using
formal methods. In this paper, we present an actor-based modeling
language with the aim to model MANETs. We address main challenges of
modeling wireless ad hoc networks such as local broadcast,
underlying topology, and its changes, and discuss how they can be
efficiently modeled at the semantic level to make their verification
amenable. The new framework abstracts the data link layer services
by providing asynchronous (local) broadcast and unicast
communication, while message delivery is in order and is guaranteed
for connected receivers. We illustrate the applicability of our
framework through two routing protocols, namely flooding and
AODVv2-11, and show how efficiently their state spaces can be
reduced by the proposed techniques. Furthermore, we demonstrate a
loop formation scenario in AODV, found by our analysis tool.

\end{abstract}

%% file: Introduction.tex
\section{Introduction\label{sec::intro}}

Applicability of wireless communications is rapidly growing from
home networks to satellite transmissions %. Also, every year lots of
%cable networks are replaced by wireless networks
due to their high accessibility and low cost. Wireless communication
has a broadcasting nature, as messages sent by each node can be
received by all nodes in its transmission range, called \emph{local
broadcast}. Therefore, by paying the cost of one transmission,
several nodes may receive the message, which leads to lower energy
consumption for the sender and throughput
improvement~\cite{cui2007distributed}.

Mobile ad hoc networks (MANETs) consist of several portable hosts
with no pre-existing infrastructure, such as routers in wired
networks or access points in managed (infrastructure) wireless
networks. In such networks, nodes can freely change their locations
so the network topology is constantly changing.  %MANET is an
%appropriate choice in critical situations like battlefields and
%disaster areas. In these environments, we need to setup a network
%very fast while requirements for deploying a planned infrastructure
%are not available due to lack of time or budget
%\cite{tseng2002broadcast}. Mobility of nodes in battlefields helps
%them to cover more area and makes them harder to be identified
%\cite{marti2000mitigating}.
For unicasting a message to a specific node beyond the transmission
range of a node, it is needed to relay the message by some
intermediate nodes to reach the desired destination. Due to lack of
any pre-designed infrastructure and global network topology
information, network functions such as routing protocols are devised
in a completely distributed manner and adaptive to topology changes. %FGHI
%In contrast to MANETs, nodes cannot change their locations in a
%Wireless Mesh Network (WMN). So,
%Wireless Mesh Networks (WMN) provide flexibility in terms of
%mobility, i.e., Mesh clients can be stationary or mobile and can
%form a client mesh network among themselves and with mesh routers
%\cite{MANETWMN}. %FGHI
%can be seen as a special case of MANETs
%due to occasion of failure of nodes or addition of new nodes.
Topology dependent behavior of wireless communication, distribution
and adaptation requirements make the design of MANET protocols
complicated and more in need of modeling and verification so that it
can be trusted. For instance, MANET protocols like the Ad hoc On Demand
Distance Vector (AODV) routing protocol \cite{perkins1999ad} has
been evolved as new failure scenarios were experienced or errors
were found in the protocol design
\cite{BhargavanOG02,LoopNamj,fehnker2013process}.

The actor model \cite{agha1985actors,hewitt1977viewing} has been
introduced for the purpose of modeling concurrent and distributed
applications. It is an agent-based language introduced by Hewitt
\cite{hewitt1977viewing}, extended by Agha to an object-based
concurrent computation model \cite{agha1985actors}. An actor model
consists of a set of actors communicating with each other through
unicasting asynchronous messages. Each computation unit, modeled by an
actor, has a unique address and mailbox. Messages sent to an actor
are stored in its mailbox. Each actor is defined through a set of
message handlers, called \textit{message servers}, to specify the
actor behavior upon processing of each message. In this model,
message delivery is guaranteed but is not in-order. This policy
implicitly abstracts away from effects of the network, i.e., 
delays over different routing paths, message conflicts, etc., and
consequently makes it a suitable modeling framework for concurrent
and distributed applications. Rebeca \cite{sirjani2004modeling} is
an actor-based modeling language which aims to bridge the gap
between formal verification techniques and the real-world software
engineering of concurrent and distributed applications. It provides
an operational interpretation of the actor model through a Java-like
syntax, which makes it easy to learn and use. Rebeca is supported by
a robust model checking tool, named Afra \cite{afra}, which takes
advantage of various reduction
techniques~\cite{jaghoori2010symmetry,sabouri2010slicing} to make
efficient verification possible. With the aim of reducing the state
space, computations, i.e., executions of message servers in actors,
are assumed to be instantaneous while message delivery is in-order.
Consequently, instructions of message servers are not interleaved
and hence, execution of message servers becomes atomic in semantic
model and each actor mailbox is modeled
through a FIFO queue. %Atomic
%execution of message servers also conforms to the fact that object
%variables are not observed during a method invocation.

In \cite{bRebeca} we introduced bRebeca as an extension to Rebeca,
to support broadcast communication which abstracts the \emph{global
broadcast communications} \cite{datanet}. To abstract the effect of
network, the order of receipts for two consequent broadcast
communications is not necessarily the same as their corresponding sends in an
actor model. Hence, each actor mailbox was modeled by a bag. The
resulting framework is suitable for modeling and efficient verification
of broadcasting protocols above the network layer, but not
appropriate for modeling MANETs in two ways: firstly the topology is
not defined, and every actor (node) can receive all messages, in
other words all nodes are connected to each other. Secondly, as
there is no topology defined, mobility is not concerned.

In this paper, we extend the actor-based modeling language bRebeca
\cite{bRebeca} to address local broadcast, topology, and its
changes. The aim of the current paper is to provide a framework to
detect malfunctions of a MANET protocol caused by conceptual
mistakes in the protocol design, rather than by an unreliable
communication. Therefore, the new framework abstracts away from the data link
layer services by providing asynchronous reliable local broadcast,
multicast, and unicast communications \cite{reliablebro,reliablemul}.
Since only one-hop communications are considered, the
message delivery is in-order and is guaranteed for connected
receivers. %%FGHI
%Our abstraction conforms to the (un-timed) abstract data
%link layer model of \cite{AbsMAC} provided to analyze MANET
%algorithms like \cite{MMB,Aleader}. Two consequent reliable
%broadcast communications are not received in-order by a common node
%in the range of their senders due to several retries made to
%guarantee that all neighbors will receive the message. The
%interleaving semantics together with out-of-order message delivery
%match the reality of concurrency in MANETs.
Consequently, each actor mailbox is modeled through a queue. % FGHI
%which makes our proposed reduction technique more effective.
The reliable communication services of the data link layer provide
feedback (to its upper layer applications) in case of (un)successful
delivery. Therefore, our framework provides \emph{conditional
unicast} to model protocol behaviors in each scenario (in the
semantic model, the status of the underlying topology defines the behavior
of actors).

The resulting framework provides a suitable means to model the behavior of
ad hoc networks in a compositional way without the need to consider %FGHI
%{\color{red}the effect of network on delivery of messages and}
asynchronous communications handled by message storages in the
computation model. However, to minimize the effect of message
storages on the growth of the state space, we exploit techniques to
reduce it. Since nodes can communicate
through broadcast and a limited form of multicast/unicast, %to an
%unspecified node,
it is possible to consider actors that have the same neighbors and local states as identical according to
the counter abstraction technique \cite{basler2009symbolic,Pnueli:2002:LCA:647771.734286,emerson1999asymmetry}. Therefore, the states whose number of actors (irrespective of their identifiers) with the same neighbors and local state are the same for each local state value, will be aggregated, thus the state space is reduced considerably. The
reduced semantics is strongly bisimilar to the original one.

To examine resistance and adaptation of MANET protocols to changes of
the underlying topology, we address mobility via arbitrary changes of
the topology at the semantic level. Since network protocols have no
control over movement of MANET nodes and mobility is an intrinsic
characteristic of such nodes, the topology should be implicitly
manipulated at the semantics. In other words, with the aim of
verifying behaviors of MANET protocols for any mobility scenario, the
underlying topology is arbitrarily changed at each semantic state.
We provide mechanisms to restrict this random changes in the topology
through specifying constraints over the topology. However, these
random changes make the state space grow exponentially while the
proposed counter abstraction technique becomes invalid. To this
end, each state is instead explored for each possible topology and
meanwhile topology information is removed from the state. Therefore,
two next states only different in their topologies are consolidated
together and hence, the state space is reduced considerably. Due to arbitrary changes of the underlying topology, states with different topologies are reachable from each other (through $\tau$-transitions denoting topology changes). We
establish that such states are branching bisimilar, and consequently a set of properties such as ACTL-X
\cite{ACTL} are preserved. The proposed reduction techniques makes
our framework scalable to verify some important properties of MANET
protocol, e.g., loop freedom, in the presence of mobility in a unified
model (cf.~generating a model for each mobility scenario).

The contributions of this paper can be summarized as follows:
\begin{itemize}
\item We extend the computation model of the actor model, in particular Rebeca, with the concepts of MANETs, i.e., asynchronous reliable
local broadcast/multicast/unicast, topology, and topology changes;
\item We apply the counter abstraction in presence of topology as a
part of semantic states to reduce state space substantially: actors
with the same neighbors, i.e., topological situations, and local states are counted together in the counter abstraction technique;
\item We show that the soundness of the counter abstraction technique is not
preserved in presence of mobility, and propose another technique to
reduce the state space.
\item We provide a tool that supports both reduction techniques and
examines invariant properties automatically. We illustrate the
scalability of our approach through the specification and verification
of two MANET protocols, namely flooding and AODV.
\item We present a complete and accurate model of the core
functionalities of a recent version of AODVv2 protocol (version 11), abstracting from its timing issues, 
and investigate its loop freedom property. We detect scenarios over
which the property is violated due to maintaining multiple unconfirmed next hops for a route 
without checking them to be loop free. We have communicated this scenario to the AODV group and they have confirmed that it can occur in practice. In response, their route information evaluation was modified, published in version $13$ of the draft.\footnote{\url{https://tools.ietf.org/html/draft-ietf-manet-aodvv2-13}} Furthermore, we verify the monotonic increase of sequence
numbers and packet delivery properties using existing model
checkers.
\end{itemize}

Our framework can also be applied to Wireless Mesh Networks (WMNs).
Unlike MANETs, WSNs have a backbone of dedicated mesh routers along
with mesh clients. Hence, they provide flexibility in terms of
mobility: in contrast to MANETs, the clients mobility has limited
effect on the overall network configuration, as the mesh routers are
fixed \cite{MANETWMN}.
%
%
% i.e. Mesh clients can be stationary or mobile and can
%form a client mesh network among themselves and with mesh routers.
%The backbone of WMNs are mesh routers which have minimal or no
%mobility and they offer network access to both mesh and conventional
%clients. WMNs aim to diversify the capabilities of ad-hoc networks.
%WMNs introduce a hierarchy in the network architecture with the
%implementation of dedicated mesh routers along with mesh clients,
%unlike the flat architecture of ad hoc networks. Another difference
%in ad hoc and WMNs is effect of mobility on network architecture. In
%ad hoc networks clients mobility completely change the network shape
%which affects the routing decisions and network performance. In case
%of WMNs the clients mobility has limited effect on the overall
%routing decisions as the mesh routers are fixed and responsible for
%routing and network configuration.
%

The paper is structured as follows. Section \ref{sec::Pre} briefly
introduces bRebeca, explain the idea behind the counter abstraction technique
and its relation to symmetry reduction technique, and explains equivalence
relations that validate our reduction techniques. Section
\ref{sec::wrebec} addresses the main modeling challenges of wireless
networks. Section \ref{sec::syntax} presents our extension to
bRebeca for modeling MANETs. In Section
\ref{sec::reduction}, we generate the state space compactly with the
aim of efficient model checking. To illustrate the applicability of
our approach, we specify the core functionalities of AODVv2-11 in Section
\ref{sec:case}. Then, in Section \ref{sec::eval}, we discuss the
efficiency of our state-space generation over two cases studies: the AODV and the flooding-based routing protocol. We illustrate our tool and possible analysis over the
models through a verification of AODV. Finally, we review some related
work in Section \ref{sec::related} before concluding the paper.

%% file: Preliminaries.tex
\section{Preliminaries}\label{sec::Pre}
\subsection{bRebeca}\label{subsec::Rebec}
Rebeca \cite{sirjani2004modeling} is an actor-based modeling
language proposed for modeling and verification of concurrent and
distributed systems. It has a Java-like syntax familiar to software
developers and it is also supported by a tool via an integrated
modeling and verification environment \cite{afra}. Due to its design
principle it is possible to extend the core language based on the
desired domain \cite{sirjani2011ten}. For example, different
extensions have been introduced in various domains such as
probabilistic systems \cite{varshosaz2012modeling}, real-time
systems \cite{aceto2011modelling}, software product lines
\cite{Sabouri}, and broadcasting environment \cite{bRebeca}. As in
this paper we intend to extend bRebeca, we briefly review its syntax
and semantics.

In bRebeca as well as in Rebeca, actors are the computation units of the system,
called rebecs  (short for reactive objects), which are instances
of the defined \emph{reactive classes} in the model.

Rebecs communicate with each other only through broadcasting message which is asynchronous. Every sent
message eventually will be received and processed by its
\emph{potential} receivers. In Rebeca, the rebecs defined as the
\emph{known rebecs} of a sender, the sender itself using the
``self'' keyword, or the sender of the message currently  processed
using the keyword ``sender" are considered as the potential
receivers. However, in bRebeca, it is assumed the network is
fully connected and therefore, all rebecs of a model constitute the
potential receivers. In other words, a broadcast message is received
by all the nodes to which a sender has a (one-hop/multi-hop) path.
So, unlike Rebeca, there is no need for declaring the known rebecs
in the reactive class definition. Due to unpredictability of
multi-hop communications, the arrival order of messages must be
considered arbitrary. Therefore, as the second difference with
Rebeca, received messages are stored in an unordered \emph{bag} in each node.% to
%model the unpredictability of arrival order of messages in such a
%network.

Every reactive class has two major parts, first the \emph{state
variables} to maintain the state of the rebec, and second the
\emph{message servers} to indicate the reactions of the rebec on
received messages. The local state of a rebec is defined in terms of
its state variables together with its message bag. Whenever a rebec
receives a message which has no corresponding message server to
respond to, it simply discards the message. Each rebec has at least one
message server called ``initial'', which acts like a constructor in
object-oriented languages and performs the initialization tasks.

A rebec is said to be \emph{enabled} if and only if it has at least
one message in its bag. The computation takes place by removing a
message from the bag % of an enabled rebec
and executing its corresponding message server atomically, after
which the rebec proceeds to process the other messages in its bag
(if any). Processing a message may have the following
consequences:
\begin{itemize}
\item it may modify the value of the state variables of the executing rebec, or
\item some messages may be broadcast to other rebecs.
\end{itemize}

Each bRebeca model consists of two parts, the
\emph{reactive classes} part and the \emph{main} part. In the
\emph{main} part the instances of the reactive classes are created
initially while their local variables are initialized.

\begin{figure}
\begin{center}
\begin{lstlisting}[language=rebeca, multicols=2]
reactiveclass MNode
{
    statevars
    {
        int my_i;
        boolean done;
    }

    msgsrv initial(int j, boolean starter)
    {
        my_i = j;
        if(starter) {
            done = true;
            send(my_i);
        } else
            done = false;
    }

    msgsrv send(int i)
    {
        if (i < my_i) {
            if (!done) {
                done = true;
                send(my_i);
            }
        } else {
            my_i = i;
            done = true;
        }
    }
}
main
{
       MNode n1(1,false);
       MNode n2(2,false);
       MNode n3(3,true);
       MNode n4(4,false);
}
\end{lstlisting}
\end{center}
\vspace{-2mm} \caption{An example in bRebeca: Max-algorithm with $4$
nodes \label{maxAlgorithm}}
\end{figure}

As an example, \figurename{~\ref{maxAlgorithm}} illustrates a simple
max finding algorithm modeled in bRebeca, referred to as
``Max-Algorithm'' \cite{maxalgorithm}. Every node in a network contains an integer value and they intend to find the maximum value
of all nodes in a distributed manner. The \texttt{initial} message server has a parameter, named \texttt{starter}. The rebec with the \texttt{starter} value  ${\it true}$ initiates the algorithm by broadcasting the first message. Whenever a node receives a
value from others, it compares this value with its current value and one of the
following scenarios happens:

\begin{itemize}
\item if it has not broadcast its value yet and its value is greater than the received one, it broadcasts its value to
others;
\item if its current value is less than the received one, it gives up broadcasting its value and updates its current value to the received
one;
\item if it has already sent its value, it only checks whether it must updates its value.
\end{itemize}

This protocol does not work on MANETs as nodes give up to rebroadcast their value after their first broadcast. The Max-Algorithm should find the maximum value among the connected nodes in MANETs. To this aim, if a node moves and connects to new nodes, it has to re-send its value as its value may be the maximum value in the currently connected nodes.

\subsection{Counter Abstraction}\label{subsec::counterAbstraction}
Since model checking is the main approach of verification in Rebeca, we need to overcome state-space
explosion, where the state space of a system grows exponentially as the number of components in the system increases. One
way to tackle this well-known problem is through applying reduction
techniques such as symmetry reduction \cite{clarke1998symmetry} and
counter abstraction \cite{basler2009symbolic,Pnueli:2002:LCA:647771.734286,emerson1999asymmetry}. Counter abstraction
is indeed a form of symmetry reduction and, in case of full symmetry, it
can be used to avoid the \emph{constructive orbit problem}, according to which finding a
unique representative of each state is NP-hard
\cite{clarke1998symmetry}. The idea of using counters and counter abstraction in model checking was first introduced in \cite{emerson1999asymmetry}. However,
the term of \emph{counter abstraction} was first presented in
\cite{Pnueli:2002:LCA:647771.734286} for the verification of
parameterized systems and further used in different studies such as
\cite{basler2009symbolic,katoen2012model}.

The idea of counter abstraction is to record the global state of a
system as a vector of counters, one per local state. Each
counter denotes the number of components currently residing in the
corresponding local state. In our work,  by``components" we mean the
actors of the system. This technique turns a model with an exponential size in $n$, i.e.\ $m^n$, into one
of a size polynomial in $n$, i.e.\ $\begin{pmatrix}
  n+m-1 \\
  m \\
\end{pmatrix}$, where $n$ and $m$ denote the number of
components and local states, respectively. Two global states $S$ and
$S'$ are considered  identical up to permutation if for every local
state $s$, the number of components residing in $s$ is the same in the
two states $S$ and $S'$, as permutation only changes the order of
elements. For example, consider a system which consists of three
components that each have only one variable $v_{i}$ of boolean type.
Three global states $({\it true},{\it true},{\it false})$, $({\it
false},{\it true},{\it true})$, and $({\it true},{\it false},{\it
true})$ are equivalent and can be abstracted into one global state
represented as $({\it true}:2,{\it false}:1)$.

\subsection{Semantic Equivalence}\label{subsec::branch}
Strong bisimilarity \cite{plotkin81} is used as a verification tool
to validate the counting abstraction reduction technique on labeled
transition systems. A labeled transition system (LTS), is defined by
the quadruple $\langle S,\rightarrow,L,s_{0} \rangle$ where $S$ is
a set of states, $\rightarrow\subseteq S\times L\times S$ a set of
transitions, $L$ a set of labels, and $s_{0}$ the initial state. Let
$s\overto{\alpha}t$ denote $(s,\alpha,t)\in\rightarrow$.

\begin{definition}[Strong Bisimilarity]\label{Def::bisim}
A binary relation $\mathcal{R}\subseteq S\times S$ is called a
strong bisimilation if and only if, for any $s_1,~s_1',~s_2$, and
$s_2'$ and $\alpha\in L$, the following transfer conditions hold:
\begin{itemize}
    \item $s_1~\mathcal{R}~s_2\wedge
    s_1\overto{\alpha}s_1'\Rightarrow(\exists s_2'\in S:s_2\overto{\alpha}s_2'\wedge s_1'~\mathcal{R}~s_2')$,
    \item $s_1~\mathcal{R}~s_2\wedge
    s_2\overto{\alpha}s_2'\Rightarrow(\exists s_1'\in S:s_1\overto{\alpha}s_1'\wedge s_1'~\mathcal{R}~s_2')$.
\end{itemize}
Two states $s$ and $t$ are called strong bisimilar, denoted by
$s\sim t$, if and only if there exists a strong bisimulation
relating $s$ and $t$.
\end{definition}

%The following theorem states that applying counter abstraction does
%preserve semantic properties of the model modulo strong
%bisimilarity.
%
%\begin{theorem}[Soundess of Counting Abstraction \cite{basler2009symbolic}]
%Let $t$ be the reduced LTS generated from $s$ using the counting
%abstraction technique. Then $s$ and $t$ are strong bisimilar.
%\end{theorem}

As explained in Section \ref{sec::intro}, mobility is addressed
through random changes of underlying topology at each semantic
state, modeled by $\tau$-transitions. We propose to
remove such transitions while the behavior of each semantic state is
explored for all possible topologies. We exploit branching
bisimilarity \cite{GlabbeekW96} to establish the reduced semantic is
branching bisimilar to the original one. Let $\overto{\tau}^*$ be
reflexive and transitive closure of $\tau$-transitions:
\begin{itemize}
\item $t\overto{\tau}^* t$;
\item $t\overto{\tau}^*s$, and $s\overto{\tau}r$, then
$t\overto{\tau}^*r$.
\end{itemize}

\begin{definition}[Branching Bisimilarity]\label{Def::brbisim}
A binary relation $\mathcal{R}\subseteq S\times S$ is called a
branching bisimilation if and only if, for any $s_1,~s_1',~s_2$, and
$s_2'$ and $\alpha\in L$, the following transfer conditions hold:
\begin{itemize}
    \item $s_1~\mathcal{R}~s_2\wedge
    s_1\overto{\alpha}s_1'\Rightarrow((\alpha=\tau \wedge s_1'~\mathcal{R}~s_2)\vee(\exists s_2',s_2''\in S:s_2\overto{\tau}^*s_2''\overto{\alpha}s_2'\wedge s_1~\mathcal{R}~s_2''\wedge s_1'~\mathcal{R}~s_2'))$,
    \item $s_1~\mathcal{R}~s_2\wedge
    s_2\overto{\alpha}s_2'\Rightarrow((\alpha=\tau \wedge s_1~\mathcal{R}~s_2')\vee(\exists s_1',s_1''\in S:s_1\overto{\tau}^*s_1''\overto{\alpha}s_1'\wedge s_1''~\mathcal{R}~s_2\wedge s_1'~\mathcal{R}~s_2'))$.
\end{itemize}
Two states $s$ and $t$ are called branching bisimilar, denoted by
$s\simeq_{br}t$, if and only if there exists a branching
bisimulation relating $s$ and $t$.
\end{definition}

%% file: BroadcastingRebecs.tex
\section{Modeling Topology and Mobility}\label{sec::wrebec}
In this section, we discuss issues brought up by extending bRebeca
to model and verify MANETs, and our
solutions to overcome these challenges. %In the following by WMNs, we
%mean networks with a stable topology, while by MANETs, we mean
%dynamic networks with a varying topology.
We assume that the
number of nodes is fixed (to make the state space finite as
explained in \cite{Param}).

\subsection{Network Topology and Mobility}
Every rebec represents a node in the MANET model. A node can
communicate only with those located in its communication range, so-called \textit{connected}. bRebeca does not define a ``topology"
concept as the network graph is considered to be connected, all
nodes are globally connected.

Mobility is the intrinsic characteristic of MANET nodes. % with usually
%exhibits the mobility behavior.
Furthermore, network protocols have no control over the movement of
MANET nodes, and hence, topology changes cannot be specified as a part of
the specification. Additionally, to verify a protocol with respect
to any mobility scenario, we need to consider all possible
topology changes while constructing the state space. To this end, we
consider the topology as a part of the states and randomly change
the underlying topology at the semantic level. To this aim, a
topology is modeled as an $n\times n $ matrix in each (global) state of the
semantic model, where $n$ is the number of nodes in the network.
Each element of this matrix, denoted by $e_{i,j}$, indicates whether
$node_i$ is \emph{connected} to $node_j$ ($e_{i,j}=1$) or not
($e_{i,j}=0$). As the communication ranges of all nodes are assumed
to be equal, connectivity is a bidirectional concept, and hence, the
resulting matrix will be symmetric. The main diagonal elements are
always $1$ to make it possible for nodes to unicast messages to themselves. (However, in the case of broadcast, our semantic rules prevent a node
from receiving its own message, see Section \ref{sec::semantics}). Changing the
topology is considered an unobservable action, modeled by a $\tau$
transition, which alters the topology matrix. Hence, each
$\tau$-transition represents a set of (bidirectional) link setups/breakdowns in the underlying topology.

To set up the initial topology of the network, the
\emph{known-rebecs} definitions, provided by the Rebeca language,
is extended to address the connectivity of rebecs.
\figurename{~\ref{Fig::Net}} shows the communication range of the
nodes in a simple network. To configure the initial topology of this
network, \emph{known-rebecs} of each rebec should be defined as
shown in \figurename{~\ref{Fig::synNet} during its instantiation
(cf.~\figurename{~\ref{maxAlgorithm}). The corresponding semantic
representation (as a part of the initial state) is shown in
\figurename{~\ref{Fig::semNet}}.

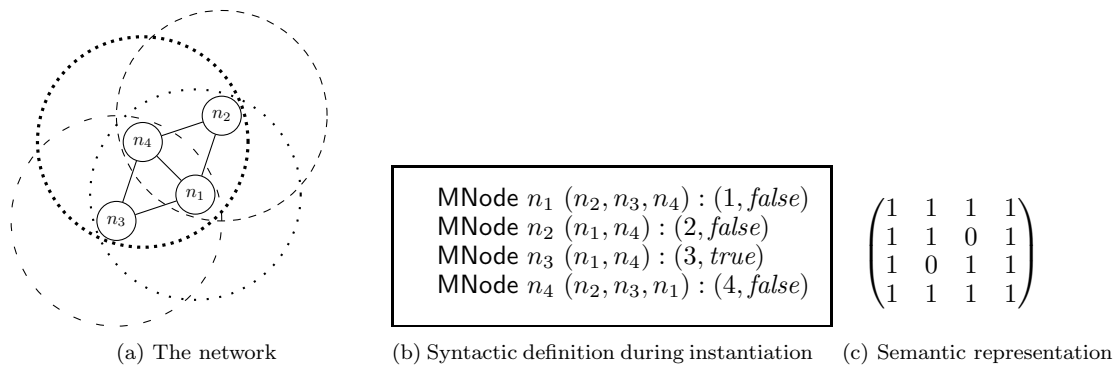
\begin{figure}
  \centering
  \begin{subfigure}[b]{0.3\textwidth}
\begin{tikzpicture}[scale=.7, transform shape]
 \node[style=circle,draw] (n3) at (1,1) {$n_3$};
 \draw [style=loosely dashed]  (1,1) circle (2) ;
  \node[style=circle,draw] (n2) at (3,3) {$n_2$};
\draw [style=dashed]  (3,3) circle (2) ;
 \node[style=circle,draw] (n4) at (1.5,2.5) {$n_4$};
\draw [style=dotted,very thick]  (1.5,2.5) circle (2) ;
\node[style=circle,draw] (n1) at (2.5,1.5) {$n_1$}; \draw
[style=loosely dotted, thick]  (2.5,1.5) circle (2) ; \draw (n3)
edge (n4); \draw (n3) edge (n1); \draw (n2) edge (n4); \draw (n2)
edge (n1); \draw (n4) edge (n1);
\end{tikzpicture}    \caption{The network}
    \label{Fig::Net}
  \end{subfigure}
    \begin{subfigure}[b]{0.34\textwidth}
    \centering
\fbox{\parbox{3cm}{\vspace{-1mm} \[\begin{array}{l}
\mathsf{MNode}~{n_1}~(n_2,n_3,n_4):(1,\mbox{\it false}) \\
\mathsf{MNode}~{n_2}~(n_1,n_4):(2,\mbox{\it false}) \\
\mathsf{MNode}~{n_3}~(n_1,n_4):(3,\mbox{\it true}) \\
\mathsf{MNode}~{n_4}~(n_2,n_3,n_1):(4,\mbox{\it false})
\end{array}\]
}}
    \caption{Syntactic definition during instantiation}
    \label{Fig::synNet}
  \end{subfigure}
  \begin{subfigure}[b]{0.25\textwidth}
\parbox{4cm}{\vspace{-2mm}
\begin{equation}\nonumber
 \begin{pmatrix}
  1 & 1 & 1 & 1 \\
  1& 1& 0 & 1 \\
  1 & 0  & 1 & 1 \\
  1 & 1& 1 & 1
 \end{pmatrix}
%\caption{modeling the topology through semantics}
\end{equation}
}
    \caption{Semantic representation}
    \label{Fig::semNet}
  \end{subfigure}
  \caption{A sample of an initial topology and its corresponding syntactic and semantic representations}
    %\label{fig:protocolExample}
\end{figure}

The connectivity matrix has $n \times n$ elements which can be either $0$ or $1$, and since on the main
diagonal we will exclusively have $1$s, we have %$(n \times n)-n$
%changeable elements which can be $1$ or $0$, and consequently 
$2^{((n\times n)-n)/2}$ possible topologies. For example, in a
network of $4$ nodes, we have $2^{(16-4)/2}=2^{6}$ possible
topologies. Considering all these topologies may lead to a state-space
explosion. Hence, we provide a mechanism to limit the possible topologies by applying some
\emph{network constraints} to characterize the set of topologies in
terms of (dis)connectivity relations to (un)pin a set of the links among
the nodes. We use the notations ${\it con}(i,j)$ or $!{\it
con}(i,j)$ to show that two nodes $i$ and $j$ are connected or
disconnected, respectively, and ${\it and}(\mathcal{C}_1,\mathcal{C}_1)$ to denote both $\mathcal{C}_1$ and $\mathcal{C}_2$ hold. For example, $!{\it con}(n_1,n_2)$
specifies that $n_1$ ($n_2$) never gets connected to $n_2$ ($n_1$),
in other words, $n_1$ never enters into $n_2$'s communication range,
and vice versa. Therefore a topology $\gamma$ is called \emph{valid} for the network
constraint $\mathcal{C}$, denoted as $\gamma \vDash \mathcal{C}$, if:
\[\begin{array}{ll}
\gamma \vDash {\it con}(i,j) \Leftrightarrow \gamma_{i,j}=1 &
\hspace{1.5cm}
 \gamma \vDash{\it and}(\mathcal{C}_1,\mathcal{C}_2) \Leftrightarrow \gamma \vDash\mathcal{C}_1\wedge \gamma
 \vDash\mathcal{C}_2\cr
\gamma \vDash !{\it con}(i,j) \Leftrightarrow \gamma_{i,j}=0 &
\hspace{1.5cm} \gamma \vDash {\it true}
\end{array}\]where $\gamma_{i,j}$ represents the element $e_{i,j}$ of the corresponding semantic model of $\gamma$, and ${\it true}$ characterizes all possible
topologies.

If the only valid topology of a network constraint is equal to the initial topology, then the underlying topology will be static. This case can be useful for modeling WMNs with stable mesh routers with no mesh clients.

\subsection{Restricted Delivery Guarantee}
The nature of communications in the wireless networks is based on
broadcast. The aim of the current paper is to provide a framework to
detect malfunctions of a MANET protocol caused by conceptual
mistakes in the protocol design, rather than by an unreliable
communication. Therefore, we consider the wireless communications in
our framework, namely local broadcast, multicast, and unicast, to be
asynchronous and reliable in order to abstract the data link layer
services. In this way, we abstract the issues related to contention
management and collision detection following the approach of
\cite{AbsMAC}. This work abstracts the services of data link
layer\footnote{Data link layer (the second layer of Open Systems
Interconnection (OSI) model) is responsible for transferring data
across the physical link. It consists of two sublayers: Logical Link
Control sublayer (LLC) and Media Access Control sublayer (MAC). LLC
is mainly responsible for multiplexing packets to their protocol
stacks identified by their IP addresses, while MAC manages accesses to the
shared media.} with the aim to design/analyze MANET protocols
irrespective to the network radio model that implements them (its
effect is captured by three delays functions). It provides reliable
local broadcast communication, with timing guarantees on the
worst-case amount of time for a message to be delivered to all its
recipients, total amount of time the sender receives its
acknowledgment, and the amount of time for a receiver to receive
some message among those currently being transmitted by its
neighbors, expressed by \textit{delay functions}. Therefore, our
approach to specify protocols relying on the abstract data link layer simplifies the study of such
protocols, and is valid as its real implementation with such reliable services exists
\cite{reliablebro,reliablemul}. In these implementations, a node can
broadcast/multicast/unicast a message successfully only to the nodes
within its communication range. Therefore, message delivery is {\it guaranteed} for the connected nodes to the sender. In the case of unicast, if the sender
is located in the receiver communication range, it will be notified,
otherwise it assumes that the transmission was unsuccessful so it can
react appropriately. Therefore, we extended bRebeca with
\textit{conditional unicast} so that it enables the model to react
accordingly based on the status of underlying topology (which
defines the delivery status in reliable communications). %However,
%broadcast/multicast does not need any delivery notification.

%Two consequent reliable broadcast communications are not received
%in-order by a common node in the range of their senders due to
%several retries made to guarantee that all neighbors will receive
%the message.
%
%Unicasting in such networks is through indicating the
%receiver of the message so that other nodes ignore it. The message
%delivery cannot be guaranteed as it depends on the proximity of the
%sender and the receiver(s). In other words,

Since we only consider one-hop communications (in contrast to the broadcast in bRebeca), the assumption about the unpredictability of multi-hop communications (with different delays) is not valid anymore, and message storages in wRebeca are modeled by queues instead of bags.

\section{wRebeca: Syntax and Semantics\label{sec::syntax}}
 In this section, we extend the syntax of bRebeca, introduced in Section
 \ref{subsec::Rebec}, with conditional unicast and multicast, topology constraint, and known rebecs to set up the initial topology. Next, we provide the semantics of wRebeca models in terms of
 LTSs.
 
 \begin{figure}
 \begin{center}
 \fbox{\parbox{\columnwidth-5mm}{\vspace{-2mm}
 % \begin{lstlisting}[captionpos=b,frame = single,  mathescape,lineskip=.05cm,tabsize=2, language=rebeca]
  \begin{align*}
  \mathrm{Model} &\Coloneqq~\mathrm{ReactiveClass}^+~\mathrm{Main}\\
  \mathrm{Main} &\Coloneqq~\mathsf{main}~\{\mathrm{RebecDecl}^+~\mathrm{ConstraintPart}~\}\\
  \mathrm{List(X)} & \Coloneqq~ \langle X,\rangle^*X ~|~\epsilon \\
  \mathrm{RebecDecl}
  &\Coloneqq~C~R~(\mathrm{List}(R))~:~(\mathrm{List}(V));\\
 \mathrm{ConstraintPart} &\Coloneqq~\mathsf{constraint}~\{\mathrm{Constraint}\}\\
 \mathrm{Constraint} &\Coloneqq~\mathrm{ConstrainDec}~|~!~\mathrm{ConstrainDec}~|~\mathsf{and}(\mathrm{Constraint}~,~\mathrm{Constraint})\\
 \mathrm{ConstrainDec} &\Coloneqq~\mathsf{con}(R~,~R)~|~~\mathsf{true}\\
  \mathrm{ReactiveClass} &\Coloneqq~\mathsf{reactiveclass}~C~\{~\mathrm{StateVars}~\mathrm{MsgServer}^*~\}\\
  \mathrm{StateVars} &\Coloneqq~\mathsf{statevars}~\{~\mathrm{VarDecl}^*~\}\\
  \mathrm{MsgServer} &\Coloneqq~\mathsf{msgsrv}~M(\mathrm{List}(T~V))~\{~\mathrm{Statement}^*~\} \\
  \mathrm{VarDecl}
  &\Coloneqq~T~V; \\
  \mathrm{Statement}
  &\Coloneqq~ \mathrm{VarDecl}~|~\mathrm{Assign}~|~\mathrm{Conditional}~|~\mathrm{Loop}~|~\mathrm{Broadcast}~|~\mathrm{Multicast}~|~\mathrm{َUnicast}~|~\mathsf{break};\\%|~\mathsf{skip}\\
  \mathrm{Assign}
  &\Coloneqq~V=Expr; \\
  \mathrm{Conditional}
  &\Coloneqq~\mathsf{if}~(Expr)~\mathrm{Block}~\mathsf{else}~\mathrm{Block}\\
  \mathrm{Block} & \Coloneqq~\mathrm{Statement}~|~\{~\mathrm{Statement}^*~\} \\
 \mathrm{Loop}&\Coloneqq~\mathsf{while}(Expr)~\mathrm{Block}\\
  \mathrm{Broadcast}
  &\Coloneqq~M(\mathrm{List}(Expr)); \\
  \mathrm{Multicast}
  &\Coloneqq~\mathsf{multicast}~(~V~,M(\mathrm{List}(Expr)));\\
 % \mathrm{Rec}
 % &\Coloneqq~\mathsf{self}~|~\mathsf{sender}~|~R\\
 \mathrm{Unicast}
 &\Coloneqq~\mathsf{unicast}~(~\mathrm{Rec}~,M(\mathrm{List}(Expr)))~\mathsf{succ:}~\mathrm{Block}~\mathsf{unsucc:}~\mathrm{Block}\\
 \mathrm{Rec} &\Coloneqq~\mathsf{self}~|~V%|~R\\
 % \mathrm{Result}
 %&\Coloneqq~;~|~\mathrm{SuccessReport}\\
 %\mathrm{SuccessReport}
 %&\Coloneqq~\mathsf{succ:}~\{~\mathrm{Statement}^*~\}~\mathsf{unsucc:}~\{~\mathrm{Statement}^*~\}\\
  % \end{lstlisting}
  \end{align*}
 }}
 \end{center}
 \vspace{-2mm} \caption{wRebeca language syntax: Angle brackets
 ($\langle~\rangle$) are used as metaparentheses. Superscript %+ is
 %used for more than one repetition, and
 * indicates zero or more
 times repetition.
  The symbols  $C$, $R$, $T$, $M$, and $V$ denote the set of classes, rebec names, types, method and variable names, respectively. The symbol $Expr$ denotes an expression, which can be an arithmetic or a boolean expression.
 \label{extendedGrammar}}
 \end{figure}

\subsection{Syntax}
The grammar of wRebeca is presented in \figurename{~\ref{extendedGrammar}}. It consists of two major parts: reactive
classes and main part. The definition of reactive classes is almost similar to
the one in bRebeca. However, the $\mathrm{main}$ part is augmented
with the $\mathrm{ConstraintPart}$, where constraints are introduced to
reduce all possible topologies in the network. The
instances of the declared reactive classes are defined in the
$\mathrm{main}$ part, before the $\mathrm{ConstraintPart}$, by indicating the name of a reactive class and an arbitrary rebec name
along with two sets of parentheses divided by the character $:$. The
first couple of parentheses is used to define the neighbors of the rebec in the 
%{\color{red}network topology,
%for static networks, and}
initial topology. %for dynamic networks. %In
%dynamic network, if no initial topology is defined or the specified
%one does not satisfy the defined constraints in
%$\mathrm{ConstraintPart}$, all topologies which satisfy the given
%constraints are considered as initial topologies.
The second couple of parentheses is used to pass values to the
initial message server. Rebecs here communicate through broadcast,
multicast, and unicast. In the broadcast statement, we simply use
the message server name along with its parameters without specifying
the receivers of a message. In contrast,  when
unicasting/multicasting a message, we also need to specify the
receiver/receivers of the message. However, there is no delivery
guarantee, depending on the location of the receiver. In case of
unicasting, the sender can react based on the delivery status. Let
$\mathsf{unicast}(\mathrm{Rec},M(\mathrm{List}(Expr)))$ indicate $\mathsf{unicast}(\mathrm{Rec},M(\mathrm{List}(Expr)))~\mathsf{succ:}\{\}~\mathsf{unsucc:}\{\}$
when the delivery status has no effect on the rebec behavior.

In addition to communication statements, there are assignment,
conditional, and loop statements. The first one is used to assign a
value to a variable. The second is used to branch based on the
evaluation of an expression: if the expression evaluates to $\it
true$, then the $\mathsf{if}$ part, and otherwise the
$\mathsf{else}$ part will be executed. Let $\mathsf{if}~(Expr)~\mathrm{Block}$ denote
$\mathsf{if}~(Expr) ~\mathrm{Block}~\mathsf{else}~\{~~\}$.
Finally, the third is used to execute a set of statements
iteratively as long as the loop condition, i.e., the boolean
expression $Expr$, holds. Furthermore, $\mathsf{break}$ can be used
to terminate its nearest enclosing loop statement and transfer the
control to the next statement. For the sake of readability, we use
$\mathsf{for}~(T~x=Expr_1;~Expr_2;~Expr_3)\{~\mathrm{Statement}^*~\}$
to denote
$T~x=Expr_1;\mathsf{while}~(Expr_2)\{~\mathrm{Statement}^*~Expr_3~\}$.
A variable can be defined in the scope of message servers as a statement similar to programming languages.

A given wRebeca model is called \textit{well-formed} if no state
variable is redefined in the scope of a message server, no two
state variables, message servers or rebec classes have identical
names, identifiers of variables, message servers and classes do not
clash, and all rebec instance accesses, message communications and
variable accesses occur over declared/specified ones and the number and type of actual
parameters correctly match the formal ones in their corresponding
message server specifications. Each $\mathsf{break}$ should occur
within a loop statement. Furthermore, the initial topology should
satisfy the network constraint and be symmetric, i.e., if $n_1$ is the
known rebec of $n_2$, then $n_2$ should be the known rebec of $n_1$.
By default, the network constraint is $\mathsf{true}$ if no network
constraint is defined, and all the nodes are disconnected if no
initial topology is defined. \vspace{4mm}

\noindent{\bf{Example:}} The flooding protocol is one of the earliest
methods used for routing in wireless networks. The flooding protocol
modeled in wRebeca is presented in \figurename{~\ref{code:flooding}}.
Every node upon receiving a packet checks whether it is the packet's
destination. If so it processes the message, otherwise it broadcasts
the message to its neighbors. To reduce the number of transferred
messages, each message contains a counter, called
\texttt{hopNum}, which shows how many times it has been
re-broadcast. If the \texttt{hopNum} is more than the specified
bound, it quits re-broadcasting.

\begin{figure}
\begin{center}
\begin{lstlisting}[language=rebeca, multicols=2]
reactiveclass Node
{
    statevars
    {
        int IP;
    }

    msgsrv initial(boolean source,int ip_)
    {
        IP=ip_;
        if(source==true)
            relay_packet(55,0,3);
    }

    msgsrv relay_packet(int data,int hopNum,int destination)
    {
        if(IP==destination)
            unicast(self, deliver_packet(data));
        else if(hopNum<3)
        {
            hopNum++;
            relay_packet(data,hopNum,destination);
        }
    }
    msgsrv deliver_packet(int data)
    {

    }
}

main
{
    Node node0 (node1):(true,0);
    Node node1 (node0,node2,node3):(false,1);
    Node node2 (node1,node3):(false,2);
    Node node3 (node1,node2):(false,3);

    constraint
    {
        and(con(node0,node1),!con(node0,node2))
    }
}
\end{lstlisting}
\end{center}
\vspace{-2mm} \caption{ Flooding protocol in a network consisting of
four nodes \label{code:flooding}}
\end{figure}

\subsection{Semantics\label{sec::semantics}}
The formal semantics of a well-formed wRebeca is expressed as an LTS. In the following, we formally
define the states, transitions, and initial states of the semantic
model generated for a given wRebeca specification. To this aim, the given specification is decomposed into
its constituent components, i.e., rebec instances, reactive classes, initial topology, and network constraint represented by the wRebeca model $\mathcal{M}$.
%{\color{red}The
%semantic model is parameterized by the boolean variable ${\it
%dynamic}$ denoting if the underlying topology is dynamic.} Hence,
The topology is implicitly changed as long as the given network
constraint is satisfied. As explained in Section \ref{sec::intro},
message server executions are atomic and their statements are not
interleaved. %{\color{red}However, their atomic execution together
%with bag policy to store received messages completely matches the
%reality of wireless networks.}
Intuitively, the global state of a wRebeca model is defined by the local
states of its rebecs and the underlying topology. Consequently, a state
transition occurs either upon atomic execution of a message server
(i.e., when a rebec processes its corresponding message in its queue), or at a random
change in the topology (modeled through unobservable
$\tau$-transitions).
% {\color{red}in case of its dynamism}.

Let $V$ denote the set of variables ranged over by $x$, and ${\it Val}$ denote
the set of all possible values for the variables, ranged over by $e$. Furthermore, we assume that the set of types $T$ consists of the integer and boolean data types, i.e., $T=\{{\it int},{\it bool}\}$.
%,
%and for the given function $f:A \rightarrow B$, ${\it dom}(f)$
%define its domain. 
We consider the default value $0\in{\it Val}$ for the integer and boolean variables since the boolean values $\it true$ and $\it false $ can be modeled by $1$ and $0$ in the semantics, respectively. The variable assignment in each scope can be
modeled by the valuation function $V\rightarrow {\it Val}$ ranged
over by $\theta$.  An assignment can be extended by writing $\theta
\cup \{y\mapsto e\}$. %(if $y\not\in {\it dom}(\theta)$) or shrunk
%$\theta \setminus \{y\mapsto -\}$ (if $y\in {\it dom}(\theta)$)
%irrespective of the value of $y$.
To monitor value assignments
regarding scope management, we specify the set of all environments
as ${\it Env}={\it Stack}(V\rightarrow {\it Val})$, ranged over by
$\upsilon$. Let ${\it upd}(\upsilon,\{y\mapsto e\})$ extend the variable
assignments of the current scope, i.e, the top of the stack, by $\{y\mapsto e\}$ if the stack is not empty. Assume ${\it Stack}()$ denotes an
empty environment. By entering into a scope, the environment
$\upsilon$ is updated by ${\it push}(\theta,\upsilon)$ where $\theta$ is empty if the scope belongs to a block (which will be extended by the declarations in the block). 
%, or can be non-empty if the scope belongs to the body of a function. 
Upon exiting from the scope, it is updated by ${\it pop}(\upsilon)$ which removes the
top of the stack. Let ${\it eval}({\it expr},\upsilon)$ denote the
value of the expression ${\it expr}$ in the context of environment
$\upsilon$, and $\upsilon[x:=e]$ the environment identical to
$\upsilon$ except that $x$ is assigned to $e$.

Assume ${\it Seq}(D)$ denotes the set of all sequences of elements
in $D$; we use notations $\langle d_1\ldots d_n\rangle$  and
$\epsilon$ for a non-empty and empty sequence, respectively. Note
that the elements in a sequence may be repeated. A FIFO queue of
elements of $D$ can be viewed as a ${\it Seq}(D)$. For instance,
$\langle2~3~2~4 \rangle\in{\it Seq}(\mathbb{N})$ denotes a FIFO queue of natural numbers where its head is $2$. For a given FIFO queue $f:{\it Seq}(D)$, assume
$f\triangleright d $ denotes the sequence obtained by appending $d$ to the end of $f$, while $d \triangleright f$ denotes the sequence with head $d$ and tail $f$. %, and $f-d$ removes first occurrence of $d$ from the
%sequence (if exists). These operators can be easily lifted to
%(multi)sets, i.e.,\ $b\oplus d_s $ and $b \ominus d_s$, where
%$d_s\subseteq D$. Two bags are equal iff the number of occurrences
%of each element $d\in D$ is identical in both bags. Furthermore, we
%use the type constructor $\powerset^M(D)$ to denote multi-sets of
%elements in $D$. For the multiset $x:\powerset^M(D_1\times D_2)$,
%assume $x\downharpoonright_{d_1} =\left\{{\left\{{d_2\mid \langle
%d_1,d_2\rangle\in x}\right\}}\right\}$, where $d_1\in D$, denotes
%the multiset of items in $x$ related to $d_1$, while $\uplus$
%denotes the unions of two multisets. We abuse the notations $b\oplus
%d_s $ and $b \ominus d_s$ if $d_s\subseteq \powerset^M(D)$.

A wRebeca model is defined through a set of reactive classes, rebec
instances, an initial topology, and a network constraint. Let $C$
denote the set of all reactive classes in the model ranged over by $c$, $R$ the
set of rebec instances ranged over by $r$, and $\mathbb{C}$ the set
of network constraints ranged over by $\mathcal{C}$. Assume $\Gamma$
is the set of all possible topologies ranged over by $\gamma$. 
%{\color{red}in case of its dynamism}.
Each reactive class $c$ is
described by a tuple $c=\langle V_{c},M_{c}\rangle$, where $V_{c}$
is the set of class state variables and $M_{c}$ the set of message
types ranged over by $m$ that its instances can respond to. We
assume that for each class $c$, we have the state variable ${\it self}\in
V_c$, and $c\in M_c$ which can be seen as its constructor in
object-oriented languages. For the sake of simplicity, we assume
that messages are parameterized with one argument, so ${\it Msg}_c$,
where $M_c = {\it Val}\rightarrow {\it Msg}_c$ defines the set of
all messages that rebec instances of  the reactive class $c$ can respond
to. The formal parameter of a message can be accessed by ${\it
fm}:M_c\rightarrow V$. Let ${\it Statement}$ denote the set of
statements ranged over by $\sigma,\delta$ (we use
$\sigma^*,\delta^*$ to denote a sequence of statements), and ${\it
body}:M_c\rightarrow {\it Seq}({\it Statement})$ specify the
sequence of statements executed by a message server. A block, denoted by $\beta$, is either defined by a statement or a sequence of statements surrounded by braces.  

A rebec instance $r$ is specified by the tuple $\langle
c,e_0\rangle$ where $c\in C$ is its reactive class, and $e_0$ defines the
value passed to the message $c$ which is initially put in the
rebec's queue. We assume a unique identifier is assigned to each
rebec instance. Let $I=\{1\ldots n\}$ denote a finite set of all
rebec identifiers ranged over by $i$ and $j$. Furthermore,  %${\it class}(i)$ denote
%the reactive class of the rebec with identifier $i$.
we use $r_i$ to denote the rebec instance $r$ with the assigned
identifier $i$.
 As explained in Section
\ref{subsec::Rebec}, a rebec in wRebeca, like Rebeca, holds its
received messages in a FIFO queue (unlike bRebeca, in which messages are
maintained in a bag).
%Each received message $m(e)$
% together with its
%sender identifier
%is stored in the rebec FIFO.

All rebecs of the model form a closed model,
denoted by $\mathcal{M}=\langle \lVert_{i \in I}r_i, C,
\gamma_0,\mathcal{C}\rangle$, where $r_i=\langle c,e_0^i\rangle$ for
some $c\in C$ and $\mathcal{C}\in \mathbb{C}$. By default,
$\mathcal{C}={\it true}$ and $\forall i,j\le n(\gamma_{0_{i,i}}=1
\wedge (i\neq j \Rightarrow \gamma_{0_{i,j}}=0))$ if no network constraint
and initial topology were defined. The (global) state of the
$\mathcal{M}$ is defined in terms of rebec's local states and the
underlying
topology. %Let $\mathit{edge}(i,j,\gamma)\rightarrow \{{\it
%true},{\it false}\}$ indicate whether the node $i$ is connected to
%node $j$ based on network topology $\gamma$: ${\it
%edge}(i,j,\gamma)=(\gamma_{i,j}==1)$.

\begin{definition}\label{Def::semantics}
The semantics of a wRebeca model $\mathcal{M}=\langle \lVert_{i \in I}r_i,C,
\gamma_0, \mathcal{C}\rangle$  is expressed by
the LTS $\langle S,L,\rightarrow,s_{0} \rangle$ where
\begin{itemize}
    \item $S \subseteq S_1\times \ldots\times S_n\times \Gamma$ is the set of global states such that $(
s_1,\ldots,s_n, \gamma)\in S$ iff $\gamma\vDash\mathcal{C}$, and
$S_i=\mathit{Env}\times\mathit{FIFO}_i$ is the set of local states
of rebec $r_i=\langle c, e_0^i\rangle$ where ${\it FIFO}_i={\it
Seq}({\it Msg}_c)$ models a FIFO queue of messages sent to the rebec $r_i$.
%{\color{red}paired by their sender identifiers.}
Therefore, each $s_i$ can be denoted by the pair $(\nu_i,f_i)$.
We use the dot notations $s_i.\nu$ and $s_i.f$ to access the
environment and FIFO queue of the rebec $i$, respectively.
    \item $L = {\it
Act}\cup\{\tau\}$ is the set of labels, where ${\it Act} = \bigcup_{c\in C}{\it Msg}_c $ ;
    \item The transition relation
${\rightarrow\subseteq S \times L \times S}$ is the least relation
satisfying the semantic rules in Table \ref{Tab::SOS};
\begin{table}
%    \centering
    \caption{wRebeca natural semantic rules}\label{Tab::SOS}
    \begin{tabular}{ll}
        \hline
        %${\it Skip}$ & \Axiom{ $\nu_i,f_1,\ldots,f_n,skip\leadsto_\gamma \nu_i,f_1,\ldots,f_n,\top$}\vspace*{2mm}\\
        ${\it Term}$: & \FAxiom{${\nu_i,f_1,\ldots,f_n,\epsilon \leadsto_\gamma \nu_i,f_1,\ldots,f_n,\top}$}\vspace*{4mm}\\
         ${\it Assign}$: & \FAxiom{ $\nu_i,f_1,\ldots,f_n,x:={\it expr};\leadsto_\gamma \nu_i[x:={\it eval}({\it
         expr},\nu_i)],f_1,\ldots,f_n,\top$}\vspace*{4mm}\\
         ${\it VDecl}$: & \FAxiom{ $\nu_i,f_1,\ldots,f_n,T~x;\leadsto_\gamma {\it upd}(\nu_i,\{x\mapsto 0\}),f_1,\ldots,f_n,\top$}\vspace*{4mm}\\
         ${\it Block}$: & $\Rule{{\it push}(\emptyset,\nu_i),f_1,\ldots,f_n,\sigma^*\leadsto_\gamma\nu_i',f_1',\ldots,f_n',\zeta}{\nu_i,f_1,\ldots,f_n,\{\sigma^*\}\leadsto_\gamma {\it pop}(\nu_i'),f_1',\ldots,f_n',\zeta}$\vspace*{4mm}\\
        ${\it Cond}_1$: & $\Rule{{\it eval}({\it expr},\nu_i)={\it true}~~~~\nu_i,f_1,\ldots,f_n,\beta_1\leadsto_\gamma \nu_i',f_1',\ldots,f_n',\zeta}{\nu_i,f_1,\ldots,f_n,{\it if}~{\it expr}~\beta_1~{\it else}~ \beta_2\leadsto_\gamma \nu_i',f_1',\ldots,f_n',\zeta}$\vspace*{4mm}\\
        ${\it Cond}_2$: & $\Rule{{\it eval}({\it expr},\nu_i)={\it false}~~~~\nu_i,f_1,\ldots,f_n,\beta_2\leadsto_\gamma \nu_i',f_1',\ldots,f_n',\zeta}{\nu_i,f_1,\ldots,f_n,{\it if}~{\it expr}~\beta_1~ {\it else}~ \beta_2\leadsto_\gamma \nu_i',f_1',\ldots,f_n',\zeta}$\vspace*{4mm}\\
        ${\it Loop}_1$: & $\Rule{\begin{array}{c}
            {\it eval}({\it expr},\nu_i)={\it true}\\\nu_i,f_1,\ldots,f_n,\beta\leadsto_\gamma \nu_i',f_1',\ldots,f_n',\top\\
            \nu_i',f_1',\ldots,f_n',{\it while}({\it expr})~\beta \leadsto_\gamma \nu_i'',f_1'',\ldots,f_n'',\top
            \end{array}}{\nu_i,f_1,\ldots,f_n,{\it while}({\it expr})~\beta \leadsto_\gamma \nu_i'',f_1'',\ldots,f_n'',\top}$\vspace*{4mm}\\
        ${\it Loop}_2$: & $\Rule{\begin{array}{c}
            {\it eval}({\it expr},\nu_i)={\it true}
            \\\nu_i,f_1,\ldots,f_n,\beta\leadsto_\gamma
            \nu_i',f_1',\ldots,f_n',\bot
            \end{array}}{\nu_i,f_1,\ldots,f_n,{\it while}({\it expr})~\beta \leadsto_\gamma \nu_i',f_1',\ldots,f_n',\top}$\vspace*{4mm}\\
        ${\it Loop}_3$: & $\Rule{
            {\it eval}({\it expr},\nu_i)={\it false}}{\nu_i,f_1,\ldots,f_n,{\it while}({\it expr})~\beta \leadsto_\gamma \nu_i,f_1,\ldots,f_n,\top}$\vspace*{4mm}\\
        ${\it BCast}$: & \FAxiom{$\nu_i,f_1,\ldots,f_n,m({\it expr});\leadsto_\gamma \nu_i,f_1',\ldots,f_n',\top$}, where $\forall k\le n(k\neq i\wedge (\gamma_{i,k}==1) \Rightarrow$\\
        & \hspace*{8cm}$[f_k'= f_k\triangleright m({\it eval}({\it expr},v_i))] [f_k'=f_k])$\vspace*{2mm}\\
        ${\it MCast}$: & \FAxiom{$\nu_i,f_1,\ldots,f_n,{\it multicast}({\it
rcvs},{\it expr});\leadsto_\gamma
        \nu_i,f_1',\ldots,f_n',\top$}, where $\forall k\le n(k\in {\it rcvs}\wedge
    (\gamma_{i,k}==1)
    \Rightarrow$\\ & \hspace*{10cm}$[f_k'=m({\it eval}({\it expr},v_i))\triangleright
    f_k][f_k'=f_k])$\vspace*{4mm}\\
       ${\it UCast}_1$: & $\Rule{\begin{array}{c}(\gamma_{i,j}==1) \\ f_j'= f_j \triangleright m({\it eval}({\it expr},v_i))\wedge \forall k\neq j(f_k'=f_k)\\\nu_i,f_1',\ldots,f_n',\beta_1\leadsto_\gamma \nu_i',f_1'',\ldots,f_n'',\zeta\end{array}}{\nu_i,f_1,\ldots,f_n,{\it unicast}(j,m({\it expr}))~{\it succ}:\beta_1~ {\it unsucc}: \beta_2\leadsto_\gamma \nu_i',f_1'',\ldots,f_n'',\zeta}$
       \vspace*{4mm}\\
        ${\it UCast}_2$: & $\Rule{(\gamma_{i,j}==0)~~~~\nu_i,f_1,\ldots,f_n,\beta_2\leadsto_\gamma \nu_i',f_1',\ldots,f_n',\zeta}{\nu_i,f_1,\ldots,f_n,{\it unicast}(j,m({\it expr}))~{\it succ}:\beta_1~ {\it unsucc}: \beta_2\leadsto_\gamma \nu_i',f_1',\ldots,f_n',\zeta}$\vspace*{2mm}\\
        %$\nu_i,f_1,\ldots,f_n,\{\sigma\}\leadsto \nu_i,f_1,\ldots,f_n,\sigma$\\
        ${\it Seq}_1$:&$\Rule{\nu_i,f_1,\ldots,f_n,\sigma_1\leadsto_\gamma \nu_i',f_1',\ldots,f_n',\top ~~~~\nu_i',f_1',\ldots,f_n',\sigma_2^*\leadsto_\gamma \nu_i'',f_1'',\ldots,f_n'',\zeta}{\nu_i,f_1,\ldots,f_n,\sigma_1 \sigma_2^* \leadsto_\gamma \nu_i'',f_1'',\ldots,f_n'',\zeta}$\vspace*{4mm}\\
        ${\it Seq}_2$: & \FAxiom{${\nu_i,f_1,\ldots,f_n,{\it break};~\sigma^* \leadsto_\gamma \nu_i,f_1,\ldots,f_n,\bot}$}
        \vspace*{4mm} \\
        ${\it Handle}$: & $\Rule{\begin{array}{c}s_i.f = m(e)\triangleright f_i  \wedge \forall k\neq i(f_k=s_k.f)\\\nu_i={\it push}(\{{\it fm}(m)\mapsto e\},s_i.\nu) \\
        \nu_i,f_1,\ldots,f_n,{\it body}(m)\leadsto_\gamma~\nu_i',f_1',\ldots,f_n',\top\end{array}}{( s_1,\ldots,s_n,\gamma)\overto{m(e)}(
        s_1',\ldots,s_n',\gamma)}$, where $\forall k\neq i(s_k'=( s_k.\nu,f_k'))\wedge s_i'=( {\it pop}(\nu_i'),f_i')$
        \vspace*{4mm} \\
        ${\it Mov}$&\FAxiom{$( s_1,\ldots,s_n,\gamma)\overto{\tau}( s_1,\ldots,s_n,\gamma')$}, where $\gamma'\models \mathcal{C}$ \vspace*{4mm}\\
        \hline
    \end{tabular}
\end{table}
    \item $s_0$ is the initial state which is defined by the combination of
initial states of rebecs and the initial topology, i.e., $s_0=\{(
s_{0}^1,\ldots,s_0^n, \gamma_0)\}$, where for the rebec $r_i=\langle
c,e_0^i\rangle$, $s_0^i=( {\it push}(\theta_0,{\it stack}()),\langle
c(e_0^i) \rangle)$ which denotes that the class variables (i.e.,
$V_c$) are initialized to the default value, denoted by $\theta_0$, and its queue includes only the
message $c(e_0^i)$, and $\gamma_0\vDash\mathcal{C}$.
\end{itemize}
\end{definition}

To describe the semantics of transitions in wRebeca in Table
\ref{Tab::SOS}, we exploit an auxiliary transition relation
$\leadsto_\gamma\subseteq( {{\it Env}\times {\it FIFO}_1\times
\ldots\times {\it FIFO}_n \times {\it Seq}({\it
Statement}))\rightarrow ({\it Env}\times {\it FIFO}_1\times
\ldots\times {\it FIFO}_n \times \{\top,\bot\}})$ to address the
effect of statement executions on the given environment of the rebec
(which executes the statements) and the queue of all rebecs.
%the effect of the given statements on the given environment and
%queues is specified by rules ${\it Skip}$, ${\it Ass}$, ${\it
%Cond}_{1,2}$, ${\it Loop}_{1-3}$, ${\it BCast}$, ${\it
%UCast}_{1,2}$, ${\it MCast}$, and ${\it Seq}_{1,2}$ and then either
Upon execution, the statements are either successfully terminated,
denoted by $\top$, or abnormally terminated, denoted by $\bot$. Let $\zeta$ range over $\{\top,\bot\}$. Rule
$\it Term$ explains that an empty statement terminates successfully.
The effect of an assignment statement, i.e., $x:={\it expr};$, is that
the value of variable $x$ is updated by ${\it eval}({\it
expr},\nu_i)$ in $\nu_i$ as explained by the rule $\it Assign$. The
variable declaration $T~x;$ extends the variable valuation corresponding to the current scope by the value assignment
$x\mapsto 0$, where $0$ is the default value for the types of $T$, as
explained in the rule $\it VDecl$. The behavior of a block is expressed by the rule $\it Block$, based on the behavior of the statements (in its scope) on the environment ${\it push}(\emptyset,\nu_i)$, where the empty valuation
function may be extended by the declarations in the scope (by rule $\it VDecl$). Thereafter, to find the effect of the block, the last scope is popped from the environment. Rules ${\it Cond}_{1,2}$ specify the
effect of the ${\it if}$ statement: If ${\it eval}({\it expr},\nu_i)$
evaluates to ${\it true}$, its effect is defined by the effect of
executing the ${\it if}$ part, otherwise the ${\it else}$ part. Rules ${\it
Loop}_{1-3}$ explain the effect of the ${\it while}$ statement; If the
loop condition evaluates to $\it true$, the effect of the ${\it
while}$ statement is defined in terms of the effect of its body by the rules
${\it Loop}_{1,2}$, otherwise it terminates immediately as specified
by the rule ${\it Loop}_3$. If the body of the ${\it while}$ statement
terminates successfully, the effect of the ${\it while}$ statement is
defined in terms of the effect of the ${\it while}$ statement on the
resulting environment and queues of its body execution as explained
by ${\it Loop}_1$. Rule ${\it Loop}_2$ expresses that if the body of
the ${\it while}$ statement terminates abnormally (due to a ${\it
break}$ statement) while its condition evaluates to $\it true$, then
it terminates successfully while taking the effect of its body
execution into account. The effect of a sequence of statements is
specified by the rules ${\it Seq}_{1,2}$. Upon successful execution of a
statement, the effect of its next statements is considered (rule ${\it Seq}_1$). A ${\it break}$ statement makes
all its next statements be abandoned (rule ${\it Seq}_2$).

The expression $b\Rightarrow [C_1][C_2]$ in the  post-conditions of rules $\it BCast$ and $MCast$ abbreviates $(b\Rightarrow C_1)\wedge (\neg b\Rightarrow C_2)$. The effects of broadcast and multi-cast communications are specified
by the rules $\it BCast$ and $\it MCast$, respectively: the message
$m({\it eval}({\it expr},\nu_i))$ is appended to the queue of all
connected nodes to the sender in case of broadcast, and all connected
nodes among the specified receivers (i.e., $\it rcvs$) in case of
multi-cast. Rules ${\it UCast}_{1,2}$ express the effect of unicast
communication upon its delivery status. If the communication was
successful (i.e., the sender was connected to the receiver), the
message is appended to the queue of the receiver while the effect of the
${\it succ}$ part is also considered (rule ${\it UCast}_1$),
otherwise only the effect of the ${\it unsucc}$ part is considered (rule
${\it UCast}_2$).

The rule $\it Handle$ expresses that the execution of a wRebeca model progresses when a rebec processes the first message of
its queue. In this rule, the message $m(e)$ is processed by the
rebec $r_i$ as $s_i.f = m(e)\triangleright f_i $. To process this
message, its corresponding message server, i.e.\ ${\it body}(m)$ is
executed. The effect of its execution is captured by the transition
relation $\leadsto_\gamma$ on the environment of $r_i$, updated by the variable assignment $\{{\it fm}\mapsto e\}$ for the scope of the message server of $m$, and the queue of all rebecs while message $m(e)$ is removed from the queue
of $r_i$. Finally, the rule ${\it Mov}$ specifies that the underlying topology is
implicitly changed at the semantic level, and the new topology
satisfies $\mathcal{C}$.
\vspace{4mm}

\noindent{\bf{Example:}} Consider the global state $(s_0,s_1,s_2,s_3,\gamma)$ such that {\small $s_0=((\push{{\it IP}\mapsto 0},\langle {\it relay\_packet}(55,0,3)\rangle)$}, {\small $s_1=(\push{{\it IP}\mapsto 1},\epsilon)$}, {\small $s_2=(\push{{\it IP}\mapsto 2},\epsilon)$}, {\small $s_3=(\push{{\it IP}\mapsto 3},\epsilon)$}, and {\small $\gamma:\begin{pmatrix}
1 & 1 & 0 & 0 \\
1& 1& 1 & 1 \\
0 & 1  & 1 &0 \\
0 & 1 & 0 & 1
\end{pmatrix})$} for the wRebeca model in \figurename{~\ref{code:flooding}} where $\push{{\it IP}\mapsto i}$ denotes ${\it push}(\{{\it IP}\mapsto i\},{\it Stack}())$. Regarding our rules, the following transition is derived:
%backup
%{\small
%	\begin{prooftree}
%		\AxiomC{${\it eval}({\it IP}=={\it des},\nu_0)$}
%		\AxiomC{${\it eval}({\it hopNum}<3,\nu_1)$}
%		\AxiomC{$\nu_2,\epsilon,\epsilon,\epsilon,\epsilon,{\it
%				hopNum}++\leadsto_\gamma
%			\nu_3,\epsilon,\epsilon,\epsilon,\epsilon,\top$}
%		\AxiomC{$\nu_3,\epsilon,\epsilon,\epsilon,\epsilon,{\it rel}({\it
%				data},\ldots)\leadsto_\gamma
%			\nu_3,\epsilon,\langle{\it
%				rel}(55,1,3)\rangle,\epsilon,\epsilon,\top$}
%		\BinaryInfC{$\nu_2,\epsilon,\epsilon,\epsilon,\epsilon,{\it
%				hopNum}++;{\it rel}({\it data},\ldots)\leadsto_\gamma \nu_3,\epsilon,\langle{\it
%				rel}(55,1,3)\rangle,\epsilon,\epsilon,\top$}
%		\BinaryInfC{$\nu_1,\epsilon,\epsilon,\epsilon,\epsilon,{\it if}({\it
%				hopNum}<3)\ldots\leadsto_\gamma \nu_4,\epsilon,\langle{\it
%				rel}(55,1,3)\rangle,\epsilon,\epsilon,\top$}
%		\BinaryInfC{$\nu_0,\epsilon,\epsilon,\epsilon,\epsilon,{\it if}({\it IP}==\ldots\leadsto_\gamma \nu_0',\epsilon,\langle{\it
%				rel}(55,1,3)\rangle,\epsilon,\epsilon,\top$}
%		\UnaryInfC{$(s_0,s_1,s_2,s_3,\gamma)\overto{{\it
%					rel}(55,0,3)}(s_0',s_1',s_2,s_3,\gamma)$}
%	\end{prooftree}}

{\small
	\begin{prooftree}
		\AxiomC{$\nu_2,\epsilon,\epsilon,\epsilon,\epsilon,{\it
				hopNum}++\leadsto_\gamma
			\nu_3,\epsilon,\epsilon,\epsilon,\epsilon,\top$}
		\AxiomC{$\nu_3,\epsilon,\epsilon,\epsilon,\epsilon,{\it rel}({\it
				data},\ldots)\leadsto_\gamma
			\nu_3,\epsilon,\langle{\it
				rel}(55,1,3)\rangle,\epsilon,\epsilon,\top$}
		\RightLabel{\scriptsize ${\it Seq}_1$}
		\BinaryInfC{$\nu_2,\epsilon,\epsilon,\epsilon,\epsilon,{\it
				hopNum}++;{\it rel}({\it data},\ldots)\leadsto_\gamma \nu_3,\epsilon,\langle{\it
				rel}(55,1,3)\rangle,\epsilon,\epsilon,\top$}
		\RightLabel{\scriptsize ${\it Block}$}
\UnaryInfC{$\nu_1,\epsilon,\epsilon,\epsilon,\epsilon,\{{\it
		hopNum}++;{\it rel}({\it data},\ldots)\}\leadsto_\gamma \nu_4,\epsilon,\langle{\it
		rel}(55,1,3)\rangle,\epsilon,\epsilon,\top~~:(\ast)$}
	\end{prooftree}}
The following inference tree uses the result of the first tree, denoted by $(\ast)$, as a part of its premise to derive the transition.
%The result of the first tree, denoted by $(\ast)$, is used as a premise in the second tree. 
{\small
\begin{prooftree}
\AxiomC{${\it eval}({\it IP}=={\it des},\nu_1)={\it false}$}
\AxiomC{${\it eval}({\it hopNum}<3,\nu_1)={\it true}$}
\AxiomC{$(\ast)$}
\RightLabel{\scriptsize ${\it Cond}_1$}
\BinaryInfC{$\nu_1,\epsilon,\epsilon,\epsilon,\epsilon,{\it if}({\it
hopNum}<3)\ldots\leadsto_\gamma \nu_4,\epsilon,\langle{\it
rel}(55,1,3)\rangle,\epsilon,\epsilon,\top$}
\RightLabel{\scriptsize ${\it Cond}_2$}
\BinaryInfC{$\nu_1,\epsilon,\epsilon,\epsilon,\epsilon,{\it if}~({\it IP}==\ldots\leadsto_\gamma \nu_1',\epsilon,\langle{\it
        rel}(55,1,3)\rangle,\epsilon,\epsilon,\top$}
\RightLabel{\scriptsize ${\it Handle}$}
\UnaryInfC{$(s_0,s_1,s_2,s_3,\gamma)\overto{{\it
rel}(55,0,3)}(s_0',s_1',s_2,s_3,\gamma)$}
\end{prooftree}}
\noindent where $\nu_1 = {\it push}(\{{\it data}\mapsto 55,{\it hopNum}\mapsto 0,{\it des}\mapsto 3\},\push{{\it IP}\mapsto 0})$, $\nu_2={\it push}(\emptyset,\nu_1)$, $\nu_3=\nu_2[{\it hopNum}:=1]$, $\nu_4={\it pop}(\nu_3)$, $\nu_1'={\it pop}(\nu_4)$, $s_0'=(\push{{\it IP}\mapsto 0},\epsilon)$, and $s_1'=(\push{{\it IP}\mapsto 1},\langle{\it
	rel}(55,1,3)\rangle)$. Note that ${\it des}$
denotes \texttt{destination}, and ${\it rel}$ refers to
\texttt{relay\_packet} message. By the rule $\it Handle$, the message ${\it rel}(55,0,3)$ in the queue of ${\it node}_0$ is processed. To this aim, the body of its message server, i.e., ${\it if}~({\it IP}==\ldots$ is executed. Since ${\it eval}({\it IP}=={\it des},\nu_1)={\it false}$}, by the rule ${\it Cond}_2$, the ${\it else}$ part (i.e., ${\it if}~({\it hopNum<3})~\ldots$) is executed. Due to ${\it eval}({\it hopNum}<3,\nu_1)={\it true}$, by the rule ${\it Cond}_1$, the ${\it if}$ part is executed.

%% file: ApplyingCounterAbstraction2.tex
\section{State-Space Reduction} \label{sec::reduction}
We extend application of the counter abstraction technique to wRebeca
models when the topology is static. To this end, the local states of
rebecs and their neighborhoods are considered. Later, we inspect the
soundness of the counter abstraction technique in the presence of mobility.
As a consequence, we propose a reduction technique based on removal
of $\tau$-transitions. Recall that the topology is static when the only valid topology of the network constraint is equal to the initial topology.

\subsection{Applying Counter Abstraction}\label{sec::count}
Assume $S_c$ is the set of local states that the instances of the
reactive class $c$ can take (i.e., $S_c={\it Env}_c\times {\it
FIFO}_c$) and $I$ is the set of rebec identifiers. To apply
counter abstraction, rebecs with an identical local state and
neighbors that are \textit{topologically equivalent} are counted
together. Two nodes $i,j\in I$ are said to be topologically
equivalent, denoted by $i\approx_\gamma j$, iff $\forall k\in
I\setminus\{i,j\}(\gamma_{ik}=\gamma_{jk})$. Intuitively, two
topologically equivalent nodes have the same neighbors (except
themselves). So if either one broadcasts, the same set of nodes
(except themselves) will receive, and if they are also connected to
each other, their counterpart (that is symmetric to the sender) will
receive. Nodes in $\mathcal{N}\subseteq I$ are called topologically
equivalent iff $\forall i,j\in \mathcal{N}\,(i\approx_\gamma j)$.
This definition implies that all topologically equivalent nodes
should be either all connected to each other, or disconnected, while
they should have the same neighbors (except themselves). Therefore,
topologically equivalent nodes will affect the same nodes when
either one broadcasts. Hence, topologically equivalent nodes with an
identical local state can be aggregated. To this aim, nodes of the
underlying topology are partitioned into the maximal sets of
topologically equivalent nodes, denoted by
$\mathcal{N}_1,\ldots,\mathcal{N}_\ell$. We define the set of
\textit{distinct local states} as $S^d=\bigcup_{c\in C}S_c$, and the
set of topology equivalence classes as
$\mathbb{T}=\{\mathcal{N}_1,\ldots,\mathcal{N}_\ell\} $.
Consequently, each global state $( s_1,\ldots,s_n,\gamma)$ is
abstracted into a vector of elements $(s^d_i,\mathcal{N}_i):c_i$ where
$s^d_i\in S^d$, $\mathcal{N}_i\in\mathbb{T}$, and $c_i$ is the
number of nodes in the topology equivalence class $\mathcal{N}_i$
that reside in the very local state $s^d_i$. The reduced global
state, called \emph{abstract global state}, is presented as follows,
where $n$ and $m$ donate the number of all rebecs and distinct
local states (i.e., $m=\left\vert{S^d}\right\vert$), respectively:
  $$S= ( (s_1^d,\mathcal{N}_1):c_1,\ldots,(s_k^d,\mathcal{N}_k):c_k),~\forall i\leq k(c_i>0\wedge \mathcal{N}_i\in\mathbb{T}),\sum_{i=1}^{k}c_i=n,~k\le n $$
%such that $\nexists i,j\le k((i\neq j)\Rightarrow \mathcal{N}_i\cup \mathcal{N}_j \mbox{ be topologically equal})$.
For instance, nodes $n_1,~n_4$, and $n_2,~n_3$ in
\figurename{~\ref{Fig::Net}} have the same neighbors, so if their
state variables and queue contents are the same, then they can be
counted together.

Recall that when the underlying topology is static, a global state
may only change upon processing a message by a rebec, since in
wRebeca the bodies of message servers execute atomically. Thus, its
corresponding abstract global state may also only change upon processing
a message by a rebec.

Counting abstraction is beneficial when the reactive classes do not have a variable that will be assigned uniquely to its instances, such as  ``unique address" as a state variable. (Note that at the semantics, rebecs have identifiers which are not a part of their local states.) For example, counter
abstraction is not effective on the specification of the
\textit{flooding protocol} given in
\figurename{~\ref{code:flooding}}, since its nodes are identified
uniquely by their IP addresses, and hence their state variables can not be
collapsed. Therefore, to take benefit of this abstraction, we
revise the example in the way that nodes are not distinguished by
their IP addresses. To this aim, the \texttt{IP} variable is replaced by the
boolean variable \texttt{destination} which identifies the sink
node, while the last parameter of the $\texttt{relay\_packet}$ message server is removed. %Furthermore, the constructor is parameterized by
%\texttt{source} to indicate the starting node to flood packets.
The revised version is shown in
\figurename{~\ref{code:flooding-revised}}.

\begin{figure}
\begin{center}
\begin{lstlisting}[language=rebeca, multicols=2]
reactiveclass Node
{
    statevars
    {
        boolean destination;
    }

    msgsrv initial(boolean source,boolean dest)
    {
        destination=dest;
        if(source==true)
            relay_packet(55,1);
    }

    msgsrv relay_packet(int data,int hopNum)
    {
        if(destination==true)
            unicast(self, deliver_packet(data));
        else if(hopNum<2)
        {
            hopNum++;
            relay_packet(data,hopNum);
        }
    }
    msgsrv deliver_packet(int data)
    {

    }
}

main
{
    Node node0 (node1):(true,false);
    Node node1 (node0,node2,node3):(false,false);
    Node node2 (node1,node3):(false,false);
    Node node3 (node1,node2):(false,true);

}
\end{lstlisting}
\end{center}
\vspace{-2mm} \caption{ The revised version of the flooding protocol to
make counter abstraction applicable in a network consisting of
four nodes \label{code:flooding-revised}}
\end{figure}

The reduction takes place on-the-fly while constructing the state
space. To this end, each global state $(
s_1,\ldots,s_n,\gamma)$ is transformed into the form $(
(s_1^d,\mathcal{N}_1):n_1,(s_2^d,\mathcal{N}_2):n_2,\ldots,(s_k^d,\mathcal{N}_k):n_k)$ such that $n_i\subseteq \mathcal{N}_i$ is the set of node identifiers
that are topologically equivalent with the local state equal to $s_i^d$, where $\mathcal{N}_i\in\mathbb{T}$. This new presentation of
the global state is called \textit{transposed global state}. The
sets $n_i$ are leveraged to update the states of the potential
receivers (known by the underlying topology) when a communication
occurs. To generate the abstract global states, each transposed
global state is processed by taking an arbitrary node from
the set assigned to a distinct local state and a topology equivalence class if the distinct local state consists of a non-empty queue. The next transposed global state is computed by executing the message handler of the head message in the queue. This is repeated for all the pairs of a distinct local state and a topology equivalence class of the transposed global state. After generating all the next transposed global states of a
transposed state, the transposed state is transformed into its
corresponding abstract global state by replacing each $n_i$ by
$\left\vert{n_i}\right\vert$. A transposed global state is
processed only if its corresponding abstract global state has not
been previously computed. During state-space generation, only the
abstract global states are stored.
\figurename{~\ref{fig:transposedGlobalState}} illustrates a global
state and its corresponding transposed global state. It is assumed
that the network consists of four nodes of the reactive class with
only one state variable $i$ and message server $\it msg$. Each row in \figurename{~\ref{fig::global}} represents a local state,
i.e., valuation of the local state variable and message queue, while
each row in \figurename{~\ref{fig::transposed}} represents a distinct
local state and a set of topologically equivalent identifiers together with those nodes of the set that reside in that distinct local state. As the topology is static, it can be removed from
the abstract/transposed global states. Furthermore, each topology equivalence class of nodes can be represented by its unique representative, e.g., the one with the minimum identifier.

\begin{figure}
   \centering
   \begin{subfigure}[b]{0.4\textwidth}
\begin{tikzpicture}[scale=.8, transform shape]
\node [draw,outer sep=0,inner sep=3,minimum size=10] at (1,11) {($\begin{array}{l} (\push{i\mapsto 1},\epsilon),
    \\ (\push{i\mapsto 2},\langle{\it msg}\rangle),\\
    (\push{i\mapsto 1},\epsilon),\\
    (\push{i\mapsto 0},\epsilon),\end{array} \begin{pmatrix}
    1 & 1 & 0 & 1 \\
    1& 1& 1 & 0 \\
    0 & 1  & 1 & 1 \\
    1 & 0 & 1 & 1
    \end{pmatrix})$};
\end{tikzpicture}
\caption{Before applying counter abstraction}
      \label{fig::global}
   \end{subfigure}
     \begin{subfigure}[b]{0.4\textwidth}
\begin{tikzpicture}[scale=.8, transform shape]
\node [draw,outer sep=0,inner sep=3,minimum size=10] at (1,11)
{($\begin{array}{l} ((\push{i\mapsto 1},\epsilon),\{1,3\}):\{1,3\},
    \\((\push{i\mapsto 2},\langle{\it msg}\rangle),\{2,4\}):\{2\},\\
    ((\push{i\mapsto 0},\epsilon),\{2,4\}):\{4\}\end{array}     )$};
\end{tikzpicture}
     \caption{After applying counter abstraction}
     \label{fig::transposed}
     \end{subfigure}
   \caption{An abstract global state and its corresponding transposed global state: assume $\push{i\mapsto e}$ denotes ${\it push}(\{i\mapsto e\},{\it Stack}())$}.
\label{fig:transposedGlobalState}
 \end{figure}

The following theorem states that applying counter abstraction
preserves semantic properties of the model modulo strong
bisimilarity. To this aim, we prove that states that are counted together are strong bisimilar. For instance, the global state similar to the one in \figurename{~\ref{fig::global}} except that the distinct local states of nodes $2$ and $4$ are swapped, is mapped into the same abstract global state that corresponds to \figurename{~\ref{fig::transposed}}.

\begin{theorem}[Soundness of Counter Abstraction]\label{The::soundnessAbs}
Assume two global states $S_1$ and $S_2$ such that for all pair of $ s^d\in S^d$ and $\mathcal{N}\in \mathbb{T}$, the number of topologically equivalent nodes of $\mathcal{N}$  that have the distinct local state $s^d$ are the same in $S_1$ and $S_2$. Then they are strongly
bisimilar.
\end{theorem}

\begin{proof}
Since the topology is static, the only transitions these states have are the result of processing messages in their rebec queues. Suppose $S_1\overto{m(e)}S_1'$ since there is a node $i$ with the local state $(\nu_i,f_i)$ in the topology equivalence class $\mathcal{N}$, where $m(e)$ is the head of $f_i$ using the semantic rule $\it Handle$ in Table \ref{Tab::SOS}. Assume that $i$ belongs to the topologically equivalent nodes $\mathcal{N}_1\subseteq \mathcal{N}$, where $((\nu_i,f_i),\mathcal{N}):\mathcal{N}_1$ is an element of the transposed global state corresponding to $S_1$. Due to the assumption, there exist topologically equivalent nodes $\mathcal{N}_2\subseteq \mathcal{N}$ in $S_2$ with the distinct local state $(\nu_i,f_i)$ where $\left\vert{\mathcal{N}_1}\right\vert=\left\vert{\mathcal{N}_2}\right\vert$. We choose an arbitrary node $j$ in $\mathcal{N}_2$ and prove that it triggers the same transition as $i$. We claim that $\forall(s_k^d,\mathcal{N}')$, the number of nodes in the topology equivalence class $\mathcal{N}'$ that are a neighbor of $i$, denoted by ${\it nb}_i$, and reside in the local state $s_k^d$ is the same to the number of the nodes in the topology equivalence class $\mathcal{N}'$ that are a neighbor of $j$, denoted by ${\it nb}_j$, with the local state $s_k^d$. Assume for the arbitrary transposed global state element $(s_l^d,\mathcal{N}'')$ this does not hold, and we consider the case where ${\it nb}_i$ has more topologically equivalent nodes than ${\it nb}_j$ in $(s_l^d,\mathcal{N}'')$. As the links are bidirectional, due to the definition of abstract/transposed global states, $i$ is the neighbor of nodes in $\mathcal{N}''$. Furthermore, as the topology is the same for $S_1$ and $S_2$ and $i,j\in\mathcal{N}$, then $j$ is also the neighbor of nodes in $\mathcal{N}''$. However, due to the assumption, the number of topologically equivalent nodes of $\mathcal{N}''$ in $S_1$ and $S_2$ that have the distinct local state $s_l^d$ are the same. So there are some topologically equivalent nodes of $\mathcal{N}''$ with the local state $s_l^d$ that are not in ${\it nb}_j$, which contradicts to fact that $j$ is the neighbor of nodes in $\mathcal{N}''$.

As both $i$ and $j$ handle the same message, they execute the same message server, and consequently the effects on their own local state and their neighbors will be the same. Therefore, $S_2\overto{m(e)}S_2'$ while $\forall (s_o^d,\mathcal{N}^\ast)$ the number of topologically equivalent nodes from the equivalence class $\mathcal{N}^\ast$ in $S_2'$ that have the distinct local state $s_o^d$ is the same to $S_1'$. A similar argumentation holds when  $S_2\overto{m(e)}S_2'$ while the inequality between ${\it nb}_i$ and ${\it nb}_j$ goes the other way.%can be dealt
%with in the same way
\end{proof}

\begin{figure}
   \centering
   \begin{subfigure}[b]{0.3\textwidth}
\begin{tikzpicture}[scale=.7, transform shape]
\node[style=circle,draw] (n1) at (1,2) {$N_1$};
\node at (1,2.5) {$(\push{{\it des}\mapsto F},\epsilon)$};
\node[style=circle,draw] (n2) at (0,1) {$N_2$};
\node at (-1.2,1.7) {$(\push{{\it des}\mapsto F},\langle {\it rel}\rangle)$};
\node[style=circle,draw] (n3) at (2,1) {$N_3$};
\node at (3.2,1.7) {$(\push{{\it des}\mapsto F},\langle {\it rel}\rangle)$};
\node[style=circle,draw] (n4) at (1,0) {$N_4$};
\node at (1,-0.5) {$(\push{{\it des}\mapsto T},\epsilon)$};
\draw (n1) edge (n2);
\draw (n1) edge (n3);
\draw (n2) edge (n4);
\draw (n3) edge (n4);
%\draw (n4) edge (n1);
\end{tikzpicture}
      \caption{Topology 1}
      \label{fig:flood-top1}
   \end{subfigure}
   \begin{subfigure}[b]{0.3\textwidth}
\begin{tikzpicture}[scale=.7, transform shape]
\node[style=circle,draw] (n1) at (1,2) {$N_1$};
\node at (1,2.5) {$(\push{{\it des}\mapsto F},\epsilon)$};
\node[style=circle,draw] (n2) at (0,1) {$N_2$};
\node at (-0.8,0.4) {$(\push{{\it des}\mapsto F},\langle {\it rel}\rangle)$};
\node[style=circle,draw] (n3) at (2,1) {$N_3$};
\node at (3.2,1.7) {$(\push{{\it des}\mapsto F},\langle {\it rel}\rangle)$};
\node[style=circle,draw] (n4) at (1,0) {$N_4$};
\node at (1,-0.5) {$(\push{{\it des}\mapsto T},\epsilon)$};
\draw (n1) edge (n2);
\draw (n1) edge (n3);
%\draw (n2) edge (n4);
\draw (n3) edge (n4);
%\draw (n4) edge (n1);
\end{tikzpicture}
      \caption{Topology 2}
      \label{fig:flood-top2}
   \end{subfigure}
     \begin{subfigure}[b]{0.3\textwidth}
\begin{tikzpicture}[scale=.8, transform shape]
\node at (1,1) {$\begin{array}{l}
    \mathsf{and}(\mathsf{con}(N_1,N_2),
    \mathsf{and}(\mathsf{con}(N_1,N_3),\\
    \mathsf{and}(\mathsf{con}(N_3,N_4),
    \mathsf{and}(!\mathsf{con}(N_1,N_2),\\
    ~~~~~~!\mathsf{con}(N_2,N_3)))))
    \end{array}$};
\end{tikzpicture}
     \caption{An example of network topology constraint}
     \label{fig:constExample}
     \end{subfigure}
   \caption{Two possible topologies for the given constraint on the flooding protocol}
   \label{fig:possible-topologies}
 \end{figure}
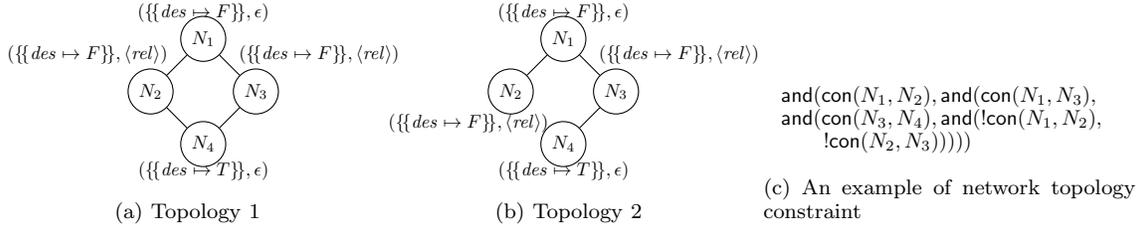

As mentioned before, the reduction is only applicable if the
network is static. This is due to the fact that if node neighborhoods may change, then
nodes which are in the same equivalence class in some state may no longer be equivalent in the next state. Consider the
flooding protocol (\figurename{~\ref{code:flooding-revised}}) for
the two topologies shown in \figurename{~\ref{fig:flood-top1}} and
\figurename{~\ref{fig:flood-top2}} (satisfying the network constraint
in \figurename{~\ref{fig:constExample}}). By applying counter
abstraction, nodes $N_2$ and $N_3$ are considered equivalent under topology $1$, but not under topology $2$.

\begin{figure}
 \centering
\begin{tikzpicture}[scale=.7, transform shape]
\node [draw,outer sep=0,inner sep=3,minimum size=10] (a1)
at (1,1)
{($\begin{array}{l}
    (\push{{\it des}\mapsto F},\epsilon),\\
    (\push{{\it des}\mapsto F},\langle {\it rel}\rangle),\\
    (\push{{\it des}\mapsto F},\langle {\it rel}\rangle),\\
    (\push{{\it des}\mapsto T},\epsilon),
    \end{array} \begin{pmatrix}
    1 & 1 & 1 & 0 \\
    1& 1& 0 & 1 \\
    1 & 0  & 1 &1 \\
    0 & 1& 1 & 1
    \end{pmatrix})$};
\node [draw,outer sep=0,inner sep=3,minimum size=10] (a2)
at (-2.5,-2) %LEFT
{($\begin{array}{l}
    (\push{{\it des}\mapsto F},\langle {\it rel}\rangle),\\
    (\push{{\it des}\mapsto F},\epsilon),\\
    (\push{{\it des}\mapsto F},\langle {\it rel}\rangle),\\
    (\push{{\it des}\mapsto T},\langle {\it rel}\rangle),
    \end{array} \begin{pmatrix}
    1 & 1 & 1 & 0 \\
    1& 1& 0 & 1 \\
    1 & 0  & 1 &1 \\
    0 & 1& 1 & 1
    \end{pmatrix})$};
\node [draw,outer sep=0,inner sep=3,minimum size=10] (a3)
at (-2.5,-5) %LEFT
{($\begin{array}{l}
    (\push{{\it des}\mapsto F},\langle {\it rel}\rangle),\\
    (\push{{\it des}\mapsto F},\epsilon),\\
    (\push{{\it des}\mapsto F},\langle {\it rel}\rangle),\\
    (\push{{\it des}\mapsto T},\langle {\it rel}\rangle),
    \end{array} \begin{pmatrix}
    1 & 1 & 1 & 0 \\
    1& 1& 0 & 0\\
    1 & 0  & 1 &1 \\
    0 & 0& 1 & 1
    \end{pmatrix})$};
\node [draw,outer sep=0,inner sep=3,minimum size=10] (a4)
at (-2.5,-8) %LEFT
{($\begin{array}{l}
    (\push{{\it des}\mapsto F},\langle {\it rel},{\it rel}\rangle),\\
    (\push{{\it des}\mapsto F},\epsilon),\\
    (\push{{\it des}\mapsto F},\epsilon),\\
    (\push{{\it des}\mapsto T},\langle {\it rel},{\it rel}\rangle),
    \end{array} \begin{pmatrix}
    1 & 1 & 1 & 0 \\
    1& 1& 0 & 0\\
    1 & 0  & 1 &1 \\
    0 & 0& 1 & 1
    \end{pmatrix})$};

\node [draw,style=dotted,very thick,outer sep=0,inner sep=3,minimum size=10]
(b2) at (4.7,-2) %RIGHT
{($\begin{array}{l}
    (\push{{\it des}\mapsto F},\langle {\it rel}\rangle),\\
    (\push{{\it des}\mapsto F},\langle {\it rel}\rangle),\\
    (\push{{\it des}\mapsto F},\epsilon),\\
    (\push{{\it des}\mapsto T},\langle {\it rel}\rangle),
    \end{array} \begin{pmatrix}
    1 & 1 & 1 & 0 \\
    1& 1& 0 & 1 \\
    1 & 0  & 1 &1 \\
    0 & 1& 1 & 1
    \end{pmatrix})$};
\node [draw,style=dotted,very thick,outer sep=0,inner sep=3,minimum size=10]
(b3) at (4.7,-5) %RIGHT
{($\begin{array}{l}
    (\push{{\it des}\mapsto F},\langle {\it rel}\rangle),\\
    (\push{{\it des}\mapsto F},\langle {\it rel}\rangle),\\
    (\push{{\it des}\mapsto F},\epsilon),\\
    (\push{{\it des}\mapsto T},\langle {\it rel}\rangle),
    \end{array} \begin{pmatrix}
    1 & 1 & 1 & 0 \\
    1& 1& 0 & 0 \\
    1 & 0  & 1 &1 \\
    0 & 0& 1 & 1
    \end{pmatrix})$};
\node [draw,style=dotted,very thick,outer sep=0,inner sep=3,minimum size=10]
(b4) at (4.7,-8) %RIGHT
{($\begin{array}{l}
    (\push{{\it des}\mapsto F},\langle {\it rel},{\it rel}\rangle),\\
    (\push{{\it des}\mapsto F},\epsilon),\\
    (\push{{\it des}\mapsto F},\epsilon),\\
    (\push{{\it des}\mapsto T},\langle {\it rel}\rangle),
    \end{array} \begin{pmatrix}
    1 & 1 & 1 & 0 \\
    1& 1& 0 & 0 \\
    1 & 0  & 1 &1 \\
    0 & 0& 1 & 1
    \end{pmatrix})$};
\draw [->,draw=black,line width=1pt] (a1) edge node[sloped,above] {$\it rel$} (a2);
\draw [->,draw=black,line width=1pt] (a2) edge node[left] {$\tau$} (a3);
\draw [->,draw=black,line width=1pt] (a3) edge node[left] {$\it rel$} (a4);
\draw [->,draw=black,line width=1pt,style=dotted,very thick] (a1) edge node[sloped,above] {$\it rel$} (b2);
\draw [->,draw=black,line width=1pt,style=dotted,very thick] (b2) edge node[left] {$\tau$} (b3);
\draw [->,draw=black,line width=1pt,style=dotted,very thick] (b3) edge node[left] {$\it rel$} (b4);
\draw [style=dashed,very thick]  (-6.2,-6.8) rectangle (8.5,-9.2);
\node [draw,outer sep=0,inner sep=3,minimum size=10] (c1)
at (-11.4,1)
{($\begin{array}{l}
    ((\push{{\it des}\mapsto F},\epsilon),\{1,4\}):\{1\},\\
    ((\push{{\it des}\mapsto F},\langle {\it rel}\rangle),\{2,3\}):\{2,3\},\\
    ((\push{{\it des}\mapsto T},\epsilon),\{1,4\}):\{4\}
    \end{array}
    )$};
\node [draw,outer sep=0,inner sep=3,minimum size=10] (c2)
at (-11.4,-2)
{($\begin{array}{l}
    ((\push{{\it des}\mapsto F},\langle {\it rel}\rangle),\{1,4\}):\{1\},\\
    ((\push{{\it des}\mapsto F},\epsilon),\{2,3\}):\{2\},\\
    ((\push{{\it des}\mapsto F},\langle {\it rel}\rangle),\{2,3\}):\{3\},\\
    ((\push{{\it des}\mapsto T},\langle {\it rel}\rangle),\{1,4\}):\{4\}
    \end{array}
    )$};
\node [draw,outer sep=0,inner sep=3,minimum size=10] (c3)
at (-11.4,-5)
{($\begin{array}{l}
    ((\push{{\it des}\mapsto F},\langle {\it rel}\rangle),\{1\}):\{1\},\\
    ((\push{{\it des}\mapsto F},\epsilon),\{2\}):\{2\},\\
    ((\push{{\it des}\mapsto F},\langle {\it rel}\rangle),\{3\}):\{3\},\\
    ((\push{{\it des}\mapsto T},\langle {\it rel}\rangle),\{4\}):\{4\}
    \end{array}
    )$};
\node [draw,outer sep=0,inner sep=3,minimum size=10] (c4)
at (-11.4,-8)
{($\begin{array}{l}
    ((\push{{\it des}\mapsto F},\langle {\it rel},{\it rel}\rangle),\{1\}):\{1\},\\
    ((\push{{\it des}\mapsto F},\epsilon),\{2\}):\{2\},\\
    ((\push{{\it des}\mapsto F},\epsilon),\{3\}):\{3\},\\
    ((\push{{\it des}\mapsto T},\langle {\it rel},{\it rel}\rangle),\{4\}):\{4\}
    \end{array}
    )$};

\draw [->,draw=black,line width=1pt] (c1) edge node[left] {$\it
rel$} (c2); \draw [->,draw=black,line width=1pt] (c2) edge
node[left] {$\tau$} (c3); \draw [->,draw=black,line width=1pt] (c3)
edge node[left] {$\it rel$} (c4); \draw[style=dashed,very thick]
(-7.8,-6.8) rectangle (-15,-9.2);
\end{tikzpicture}

 \caption{
Comparing a part of the flooding protocol's state space with/without
applying counter abstraction in a dynamic network. The two dashed bordered
states are not strong bisimilar since in the right figure there is a
global state in which only one node has two ${ \it rel}$ messages in
its queue while in the left figure there are two nodes with queues containing two ${\it rel}$ messages. Note that $T,F$ stand for ${\it true}, {\it false}$, ${\it des}$
denotes \texttt{destination}, and ${\it rel}$ refers to
\texttt{relay\_packet} messages. For simplicity the message
parameters are not shown in the figure.
 }
  \label{fig:stateSpaceFloodingCounterDynamic}
    \end{figure}
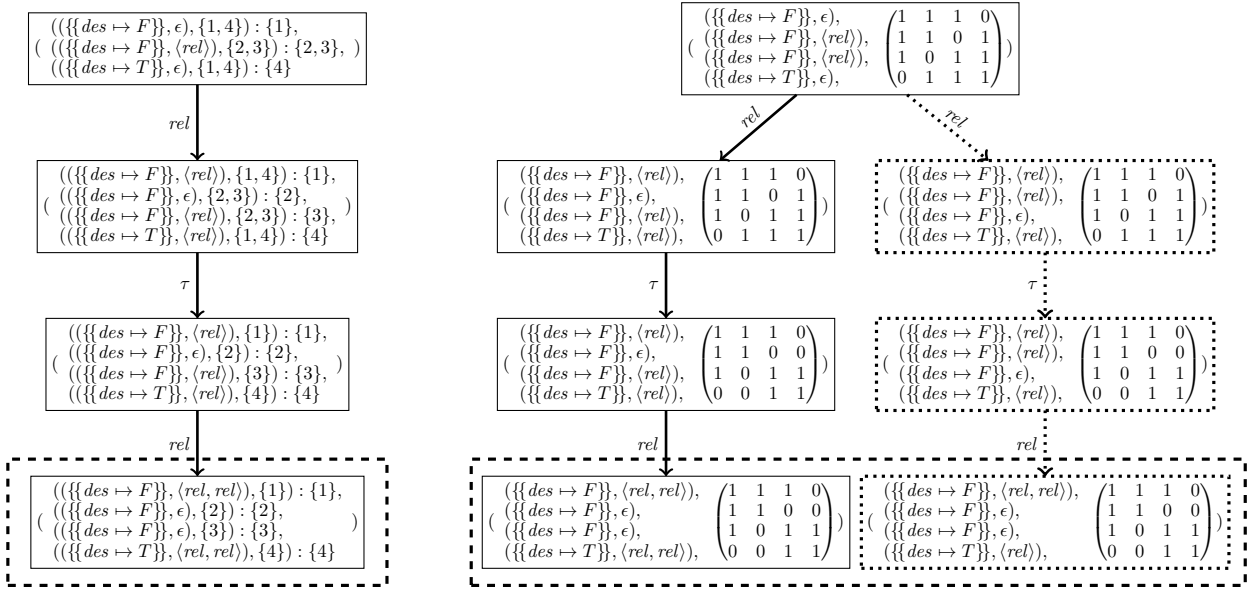

To illustrate that counter abstraction is not applicable to systems with a dynamic topology,
\figurename{~\ref{fig:stateSpaceFloodingCounterDynamic}} shows a part
of the state space of the flooding protocol with a change in the
underlying topology (from \figurename{~\ref{fig:flood-top1}} to
\figurename{~\ref{fig:flood-top2}}) with/without applying counter
abstraction, where only these two topologies are possible. As predicted, the reduced state space is not strong
bisimilar (see Section \ref{subsec::branch} for the definition) to
its original state space. During transposed global state generation, the next state is only generated for node $2$ with the distinct local state $(\push{{\it des}\mapsto F},\langle {\it rel}\rangle)$ from the equivalence class $\{2,3\}$. Therefore, it is obvious that the next states in the left LTS of \figurename{~\ref{fig:stateSpaceFloodingCounterDynamic}} can be matched to the states with the solid borders in the right LTS. However, the solid bordered states are not strong bisimilar to the dotted ones in the right LTS. As explained in Section
\ref{sec::intro}, the reduced LTS should be strong bisimilar to its
original one to preserve all properties of its original model.

To take a better advantage of the reduction technique, the message storages can be modeled as bags. However, such an abstraction results in more interleavings of messages which do not necessarily happen in reality, and hence, an effort to inspect if a given trace (of the semantic model) is a valid scenario in the reality is needed. This effort is only tolerable if the state space reduces substantially.

\subsection{Eliminating $\tau$-Transitions}
Instead of modifying the underlying topology, modeled by
$\tau$-transitions, messages can be processed with respect to all
possible topologies (not only to the current underlying topology).
Therefore, all $\tau$-transitions are eliminated and only those that correspond to processing of messages are kept. The following theorem
expresses that removal of $\tau$-transitions and topology
information from the global states preserves properties of the
original model modulo branching bisimulation, such as ACTL-X~\cite{ACTL}. In fact, by exploiting a result from~\cite{ACTL} about the correspondence between the equivalence induced by ACTL-X and branching bisimulation, the ACTL-X fragments of CACTL \cite{ghassemi2013model}, introduced to specify MANET properties, and $\mu$-calculus are also preserved. We show in Section \ref{subsec::properties} that important properties of MANET protocols can be still verified over reduced state spaces.

\begin{figure}
    \centering
    %\includegraphics[width=8cm]{visio}
%        \resizebox{#1\linewidth}{!}{
      \input{visio}
%    }
    \caption{Relation $\mathcal{R}$ matches states $
        ( s, \gamma)$ of $T_0$ to $s$ of $T_1$.}\label{Fig::branch}
\end{figure}
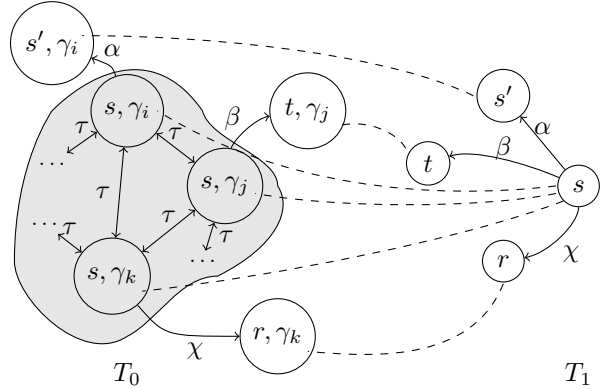

\begin{theorem}[Soundness of $\tau$-Transition Elimination]\label{The::red}
For the given LTS $T_0\equiv\langle S\times
\Gamma,\rightarrow,L,( s_{0},\gamma_0) \rangle$, assume that $( s,\gamma)\overto{\alpha}( t,\gamma')
\Rightarrow (
\gamma=\gamma')\vee(\alpha=\tau\wedge s= t)$,
and $\forall \gamma,\gamma'\in\Gamma:(
s,\gamma)\overto{\tau}( s,\gamma')$. If
$T_1\equiv\langle S,\rightarrow',L,s_{0} \rangle$, where
$\rightarrow'=\{(s,\alpha,t)\,\mid\,((s,\gamma),\alpha,(t,\gamma))\in \rightarrow\}$, then $(s_0,\gamma_0)\simeq_{br}s_0$.
\end{theorem}
\begin{proof}
Construct $\mathcal{R}=\{(( s,\gamma) , s) | s\in S,
\gamma\in\Gamma\}$ as shown in Figure \ref{Fig::branch}. We show
that $\mathcal{R}$ is a branching bisimulation. To this aim, we show
that it satisfies the transfer conditions of Definition
\ref{Def::brbisim}. For an arbitrary relation $(
s,\gamma) ~\mathcal{R}~ s$, assume $(
s,\gamma)\overto{\alpha}( t,\gamma')$. If
$\alpha=\tau$, then two cases can be distinguished: (1) either
$\gamma\neq\gamma'$, and hence by definition of $T_0$, $s=t$ holds which
concludes $( t,\gamma') ~\mathcal{R}~ s$, (2) or
$\gamma=\gamma'$ and by definition of $T_1$, $s\pverto{\alpha}t$,
and $( t,\gamma') ~\mathcal{R}~ t$. If $\alpha\neq \tau$, then by
definition of $T_0$, $\gamma=\gamma'$ and hence by definition of
$T_1$, $s\pverto{\alpha}t$, and $( t,\gamma')
~\mathcal{R}~ t$. Whenever $s\pverto{\alpha}t$, then by definition of $T_1$ there exists
$\gamma'$ such that $( s,\gamma')\overto{\alpha}(
t,\gamma')$ and hence, $( t,\gamma') ~\mathcal{R}~ t$.
%Furthermore by definition of $T_0$, $(
%s,\gamma)\overto{\tau}( s,\gamma')$ and $(
%s,\gamma') ~\mathcal{R}~ s$.
Consequently $\mathcal{R}$ is a
branching bisimulation relation.
\end{proof}

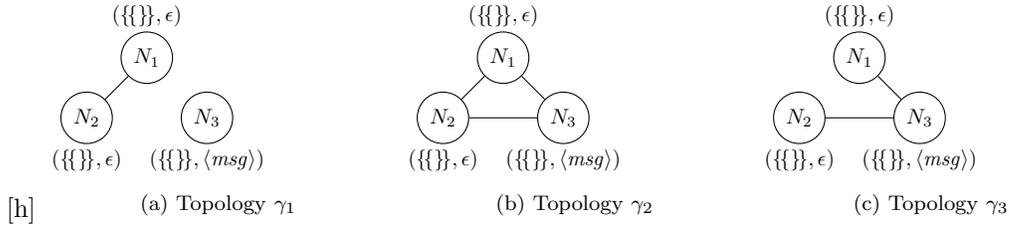
\begin{figure}[h]
  	\centering
  	\begin{subfigure}[b]{0.28\textwidth}
  		\begin{tikzpicture}[scale=.8, transform shape]
  		\node[style=circle,draw] (n1) at (1,2) {$N_1$};
  		\node at (1,2.7) {$(\push{},\epsilon)$};
  		\node[style=circle,draw] (n2) at (0,1) {$N_2$};
  		\node at (0,0.3) {$(\push{},\epsilon)$};
  		\node[style=circle,draw] (n3) at (2,1) {$N_3$};
  		\node at (2,0.3) {$(\push{},\langle {\it msg}\rangle)$};
  		\draw (n1) edge (n2);
  		%\draw (n1) edge (n3);
  		%\draw (n2) edge (n3);
  		\end{tikzpicture}
  		\caption{Topology $\gamma_1$}
  		\label{fig:TauTop1}
  	\end{subfigure}
  	\begin{subfigure}[b]{0.28\textwidth}
  		\begin{tikzpicture}[scale=.8, transform shape]
  		\node[style=circle,draw] (n1) at (1,2) {$N_1$};
  		\node at (1,2.7) {$(\push{},\epsilon)$};
  		\node[style=circle,draw] (n2) at (0,1) {$N_2$};
  		\node at (0,0.3) {$(\push{},\epsilon)$};
  		\node[style=circle,draw] (n3) at (2,1) {$N_3$};
  		\node at (2,0.3) {$(\push{},\langle {\it msg}\rangle)$};
  		\draw (n1) edge (n2);
  		\draw (n1) edge (n3);
  		\draw (n2) edge (n3);
  		\end{tikzpicture}
  		\caption{Topology $\gamma_2$}
  		\label{fig:TauTop2}
  	\end{subfigure}
  	\begin{subfigure}[b]{0.28\textwidth}
  		\begin{tikzpicture}[scale=.8, transform shape]
  		\node[style=circle,draw] (n1) at (1,2) {$N_1$};
  		\node at (1,2.7) {$(\push{},\epsilon)$};
  		\node[style=circle,draw] (n2) at (0,1) {$N_2$};
  		\node at (0,0.3) {$(\push{},\epsilon)$};
  		\node[style=circle,draw] (n3) at (2,1) {$N_3$};
  		\node at (2,0.3) {$(\push{},\langle {\it msg}\rangle)$};
  		%\draw (n1) edge (n2);
  		\draw (n1) edge (n3);
  		\draw (n2) edge (n3);
  		\end{tikzpicture}
  		\caption{Topology $\gamma_3$}
  		\label{fig:TauTop3}
  	\end{subfigure}
  	\caption{All possible topologies considered during state-space generation of \figurename{~\ref{fig:ssWirelesswithAndwitoutReduction}}}
  	\label{fig:topologiesRemovingTau}
  \end{figure}

We remark that the labeled transitions $T_0$ and $T_1$ in the Theorem \ref{The::red} specify the state space of wRebeca models before and after elimination of $\tau$-transitions, respectively. As an example, consider a network which consists of three nodes, which are the instances
of a reactive class with no state variable and only one message,
$\it msg$. The message server $\it msg$ has only one statement
to broadcast the message $\it msg$ to its neighbors. We assume that the
set of all possible topologies is restricted by a network constraint
to the three topologies depicted in
\figurename{~\ref{fig:topologiesRemovingTau}}. Consider
the global state in which only $N_3$ has one $\it msg$ in its queue.
%as
%shown in \figurename{\ref{fig:topologiesRemovingTau}}.

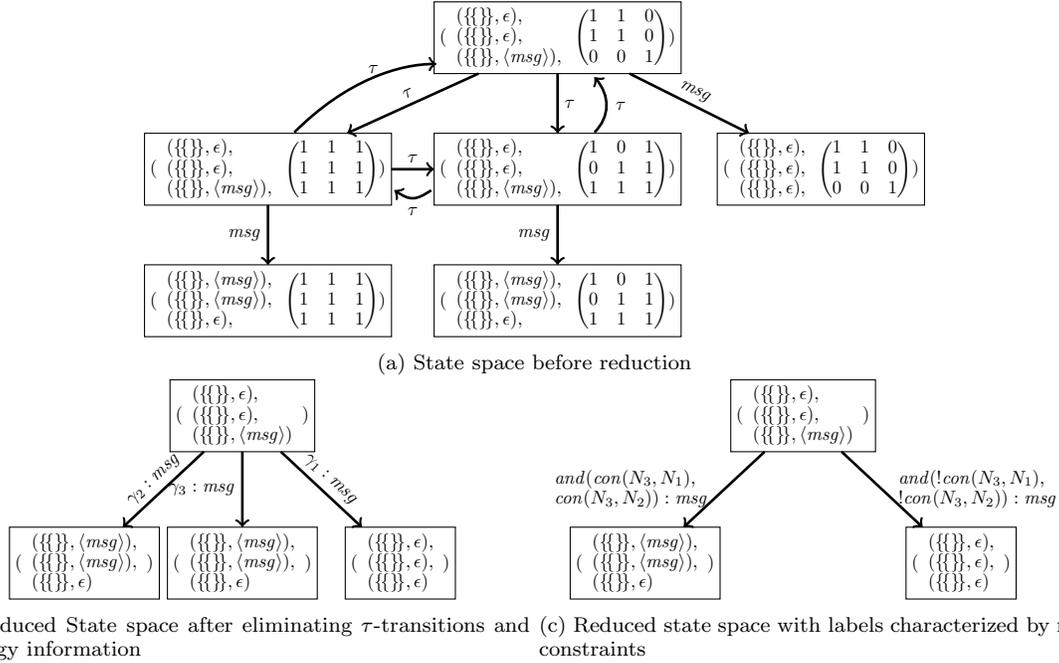
\begin{figure}
    \centering
      \begin{subfigure}[b]{\textwidth}
      \centering
    \begin{tikzpicture}[scale=0.7,transform shape]
    \node [draw,outer sep=0,inner sep=3,minimum size=10] (a1)
    at (0,1.5)
    {$(\begin{array}{l}
        (\push{},\epsilon),\\
        (\push{},\epsilon),\\
        (\push{},\langle {\it msg}\rangle),\\
        \end{array} \begin{pmatrix}
        1 & 1 & 0  \\
        1& 1& 0  \\
        0 & 0  &  1\\
        \end{pmatrix})$};
    \node [draw,outer sep=0,inner sep=3,minimum size=10] (a2)
    at (-5.5,-1)
    {$(\begin{array}{l}
        (\push{},\epsilon),\\
        (\push{},\epsilon),\\
        (\push{},\langle {\it msg}\rangle),\\
        \end{array} \begin{pmatrix}
        1 & 1 & 1  \\
        1& 1& 1  \\
        1 & 1  &  1\\
        \end{pmatrix})$};
    \node [draw,outer sep=0,inner sep=3,minimum size=10] (a3)
    at (0,-1)
    {$(\begin{array}{l}
        (\push{},\epsilon),\\
        (\push{},\epsilon),\\
        (\push{},\langle {\it msg}\rangle),\\
        \end{array} \begin{pmatrix}
        1 & 0 & 1  \\
        0& 1& 1  \\
        1 & 1  &  1\\
        \end{pmatrix})$};
    \node [draw,outer sep=0,inner sep=3,minimum size=10] (a4)
    at (5,-1)
    {$(\begin{array}{l}
        (\push{},\epsilon),\\
        (\push{},\epsilon),\\
        (\push{},\epsilon),\\
        \end{array} \begin{pmatrix}
        1 & 1 & 0  \\
        1& 1& 0  \\
        0 & 0  &  1\\
        \end{pmatrix})$};
    \node [draw,outer sep=0,inner sep=3,minimum size=10] (a5)
    at (-5.5,-3.5)
    {$(\begin{array}{l}
        (\push{},\langle {\it msg}\rangle),\\
        (\push{},\langle {\it msg}\rangle),\\
        (\push{},\epsilon),\\
        \end{array} \begin{pmatrix}
        1 & 1 & 1  \\
        1& 1& 1  \\
        1 & 1  &  1\\
        \end{pmatrix})$};
    \node [draw,outer sep=0,inner sep=3,minimum size=10] (a6)
    at (0,-3.5)
    {$(\begin{array}{l}
        (\push{},\langle {\it msg}\rangle),\\
        (\push{},\langle {\it msg}\rangle),\\
        (\push{},\epsilon),\\
        \end{array} \begin{pmatrix}
        1 & 0 & 1  \\
        0& 1& 1  \\
        1 & 1  &  1\\
        \end{pmatrix})$};
    \draw [->,draw=black,line width=1pt] (a1) edge node[sloped,above] {$\tau$} (a2);
    \draw [->,draw=black,line width=1pt] (a1) edge node[right] {$\tau$} (a3);
    \draw [->,draw=black,line width=1pt] (a1) edge node[sloped,above] {$\it msg$} (a4);

    %\draw[->,draw=black,line width=1pt] (-5,-0.3) .. controls (-4,0.5) and (-3,1)  .. (-2.7,1);
    %\draw (-7,3) .. controls (-6,5) and (-3,5) .. (-2,3);

    \draw[->,draw=black,line width=1pt] (-5,-0.3) .. controls (-4,0.7) and (-3,1) .. (-2.3,1)  ;
    \node at (-3.5,0.9) {$\tau$};
    \draw [->,draw=black,line width=1pt] (a2) edge node[left] {$\it msg$} (a5);
    \draw [->,draw=black,line width=1pt] (a2) edge node[above] {$\tau$} (a3);

    \draw [->,draw=black,line width=1pt] (-2.4,-1.4) .. controls (-2.7,-1.6) and (-2.8,-1.6) .. (-3.1,-1.4);
    \node at (-2.75,-1.8) {$\tau$};
    \draw [->,draw=black,line width=1pt] (a3) edge node[left] {$\it msg$} (a6);

    \draw[->,draw=black,line width=1pt] (0.7,-0.3) .. controls (1,0) and (1,0.4) .. (0.7,0.75);
    \node at (1.2,0.2) {$\tau$};
    \end{tikzpicture}
    \caption{
    State space before reduction
    }
    \label{fig:stateSpaceWirelessWithoutReduction}
      \end{subfigure}
       \begin{subfigure}[b]{0.48\textwidth}
        \centering
        \begin{tikzpicture}[scale=0.7,transform shape]
        \node [draw,outer sep=0,inner sep=3,minimum size=10] (a1)
        at (0,1.8)
        {$(\begin{array}{l}
            (\push{},\epsilon),\\
            (\push{},\epsilon),\\
            (\push{},\langle {\it msg}\rangle)\\
            \end{array} )$};
        \node [draw,outer sep=0,inner sep=3,minimum size=10] (a2)
        at (-3,-1)
        {$(\begin{array}{l}
            (\push{},\langle {\it msg}\rangle),\\
            (\push{},\langle {\it msg}\rangle),\\
            (\push{},\epsilon)\\
            \end{array} )$};
        \node [draw,outer sep=0,inner sep=3,minimum size=10] (a3)
        at (0,-1)
        {$(\begin{array}{l}
            (\push{},\langle {\it msg}\rangle),\\
            (\push{},\langle {\it msg}\rangle),\\
            (\push{},\epsilon)\\
            \end{array} )$};
        \node [draw,outer sep=0,inner sep=3,minimum size=10] (a4)
        at (3,-1)
        {$(\begin{array}{l}
            (\push{},\epsilon),\\
            (\push{},\epsilon),\\
            (\push{},\epsilon)\\
            \end{array} )$};
        \draw [->,draw=black,line width=1pt] (a1) edge node[sloped,above] {$\gamma_2:{\it msg}$} (a2);
        \draw [->,draw=black,line width=1pt] (a1) edge node[sloped,above] {$\gamma_1:{\it msg}$} (a4);
        \draw [->,draw=black,line width=1pt] (a1) edge node[left] {$\gamma_3:{\it msg}$} (a3);
        \end{tikzpicture}
        \caption{Reduced State space after eliminating $\tau$-transitions and topology information}
        \label{fig:stateSpaceWirelessWithoutConnection}
       \end{subfigure}
       \begin{subfigure}[b]{0.48\textwidth}

        \begin{tikzpicture}[scale=0.7,transform shape]
        \node [draw,outer sep=0,inner sep=3,minimum size=10] (a1)
        at (0,1.8)
        {$(\begin{array}{l}
            (\push{},\epsilon),\\
            (\push{},\epsilon),\\
            (\push{},\langle {\it msg}\rangle)\\
            \end{array} )$};
        \node [draw,outer sep=0,inner sep=3,minimum size=10] (a2)
        at (-3,-1)
        {$(\begin{array}{l}
            (\push{},\langle {\it msg}\rangle),\\
            (\push{},\langle {\it msg}\rangle),\\
            (\push{},\epsilon)\\
            \end{array} )$};
        \node [draw,outer sep=0,inner sep=3,minimum size=10] (a4)
        at (3,-1)
        {$(\begin{array}{l}
            (\push{},\epsilon),\\
            (\push{},\epsilon),\\
            (\push{},\epsilon)\\
            \end{array} )$};
        \draw [->,draw=black,line width=1pt] (a1) edge node[left] {$\begin{array}{l}{\it and}({\it con}(N_3,N_1),\\{\it con}(N_3,N_2)):{\it msg}\end{array}$} (a2);
        \draw [->,draw=black,line width=1pt] (a1) edge node[right] {$\begin{array}{l}{\it and}(!{\it con}(N_3,N_1),\\!{\it con}(N_3,N_2)):{\it msg}\end{array}$} (a4);
        \end{tikzpicture}

        \caption{Reduced state space with labels characterized by network constraints
        }
        \label{fig:stateSpaceWirelessWithoutConnectionCACTL}
       \end{subfigure}
      \caption{State space before and after applying reduction
      }
      \label{fig:ssWirelesswithAndwitoutReduction}
    \end{figure}

The state space of the above imaginary model before reduction is
presented in
\figurename{~\ref{fig:stateSpaceWirelessWithoutReduction}}, where
transitions take place by processing messages or changing the
topology. \figurename{~\ref{fig:stateSpaceWirelessWithoutConnection}}
illustrates the state space after eliminating $\tau$-transitions and topology information. Connectivity information is removed from the global states, as in each state its transitions are derived for all possible topologies. In
this approach, transition labels are paired with the topology to
denote the topology-dependent behavior of transitions. The two transitions labeled with $\gamma_2$ and $\gamma_3$ can be
merged by characterizing the links that make communication from
$N_3$ to $N_1$ and $N_2$; i.e., from the sender to the receivers.
Such links can be characterized by the network constraints depicted
in \figurename{~\ref{fig:stateSpaceWirelessWithoutConnectionCACTL}}.
In this model, a state is representative of all possible topologies. The resulting semantic model, called \textit{Constrained
    Labeled Transition System} (CLTS), was introduced
in \cite{FatemehTCS} as the semantic model to compactly model MANET
protocols. Another advantage of a CLTS is its model checker to verify
topology-dependant behavior of MANETs \cite{ghassemi2013model}. The
properties in wireless networks are usually pre-conditioned to existence of a path between two nodes. This model checker takes benefit
of network constraints over transitions and assures a property holds if the required
paths hold (inferred from the traversed network constraints).

%% file: visio.tex
%\usetikzlibrary{decorations.pathmorphing}
\begin{tikzpicture}
\draw[fill=black!10]  (2,1) .. controls (2.4,1.4) and (3.3,2) .. (4,1) .. controls (4.1,0.6) and (5,0.4) .. (5,0) .. controls (5.2,-0.3) and (5,-0.8) .. (4.2,-1.2) .. controls (3.7,-1.2) and (3.4,-1.8) .. (3,-2) .. controls (2.4,-2.4) and (1.5,-1.5) .. (1.5,-1) .. controls (1.5,0.3) and (2,1) .. (2,1);

\node[style=circle,draw] (s1) at (9,0) {$s$};
\node[style=circle,draw] (s2) at (8,1.2) {$s'$};
\node[style=circle,draw] (t) at (7,0.3) {$t$};
\node[style=circle,draw] (r) at (8,-1) {$r$};
\draw[->] (s1) edge node[above] {$\alpha$}(s2);
 \draw[->] (s1) .. controls (8.3,0.3) and (7.6,0.5) .. (t);
\node at (8,0.5) {$\beta$};
 \draw[->] (s1) .. controls (9,-0.5) and (8.6,-0.9) .. (r);
\node at (8.9,-0.9) {$\chi$};
\node[style=circle,draw,minimum size=10] (ss1) at (3,1) {$s,\gamma_i$};
\node[style=circle,draw,minimum size=10] (ss2) at (4.3,0) {$s,\gamma_j$};
\node[style=circle,draw,minimum size=10] (ss3) at (2.8,-1.2) {$s,\gamma_k$};
\draw[<->] (ss1) edge node[above] {$\tau$}(ss2);
\draw[<->] (ss1) edge node[left] {$\tau$}(ss3);
\draw[<->] (ss3) edge node[above] {$\tau$}(ss2);
\node[style=circle,draw,minimum size=10] (tt1) at (5.4,1) {$t,\gamma_j$};
\draw[->] (ss2) .. controls (4.4,0.6) and (4.7,0.9) .. (tt1);
\node at (4.4,0.9) {$\beta$};
\node[style=circle,draw,minimum size=10] (tt2) at (2,1.9) {$s',\gamma_i$};
\draw[->] (ss1) .. controls (2.8,1.6) .. (tt2);
\node (v1) at (2.8,1.8) {$\alpha$};
\node[style=circle,draw,minimum size=10] (tt3) at (5,-2) {$r,\gamma_k$};
\draw[->] (ss3) .. controls (3.5,-2) .. (tt3);
\node at (3.9,-2.2) {$\chi$};
\node (e1) at (4,-1) {\footnotesize $\ldots$};
\draw[<->] (ss2) edge node[right] {$\tau$}(e1);
\node (e2) at (1.9,-0.5) {\footnotesize $\ldots$};
\draw[<->] (ss3) edge node[above] {$\tau$}(e2);
\node (e3) at (2,0.3) {\footnotesize $\ldots$};
\draw[<->] (ss1) edge node[above] {$\tau$}(e3);
\draw[dashed] (8,-1.3) .. controls (7.6,-2.3) and (6.4,-2.3) .. (5.4,-2.2);
\draw[dashed] (6.7,0.5) .. controls (6.6,0.7) and (6.2,0.9) .. (5.8,0.8);
\draw[dashed] (7.6,1.2) .. controls (6.6,1.7) and (4.4,2) .. (2.4,2);
\draw[dashed]  (8.7,0) .. controls (7,-0.2) and (5,0) .. (3.4,1);
\draw[dashed]  (8.7,-0.1) .. controls (7.7,-0.3) and (5.8,-0.3) .. (4.7,-0.1);
\draw[dashed]  (8.8,-0.2) .. controls (7.5,-0.7) and (5,-1.2) .. (3.2,-1.4);
\node at (3,-2.5) {$T_0$};
\node at (9,-2.5) {$T_1$};
\end{tikzpicture}
%\draw [fill=black!20] (0,2) ..controls (0.5,2.5) and (1,2) .. (1.5,1.5);

%\draw[decorate,decoration=saw, fill=black!20]  (3,1) ellipse (1 and 2);

%% file: CaseStudies.tex
 \section{Modeling the AODVv2 Protocol \label{sec::AODV}}
\label{sec:case} To illustrate the applicability of the proposed
modeling language, the AODVv2 \footnote{\url{https://tools.ietf.org/html/draft-ietf-manet-aodvv2-11}}(i.e.,\ version 11)
protocol is modeled.
%with the aim of reducing its state space in a
%dynamic network by applying our reduction technique.
The AODV is a
popular routing protocol for wireless ad hoc networks, first
introduced in \cite{perkins1999ad}, and later revised several times.
%, in this paper we focused on the recent version 11.

In this algorithm, routes are constructed
dynamically whenever requested. Every node has its own routing table
to maintain information about the routes of the received packets.
When a node receives a packet (whether it is a route discovery or
data packet), it updates its own routing table to keep the shortest
and freshest path to the source or destination of the received packet.
Three different tables are used to store information about the neighbors, routes and received messages:
\begin{itemize}
    \item neighbor table: keeps the adjacency states of the node's neighbors. The neighbor state can be one of the following values:
        \begin{itemize}
            \item \it{Confirmed}: indicates that a bidirectional link to that neighbor exists. This state is achieved either through receiving a \it{rrep} message in response to a previously sent \it{rreq} message, or a \it{RREP\_Ack}  message as a response to a previously sent \it{rrep} message (requested an \it{RREP\_Ack}) to that neighbor.
            \item \it{Unknown}: indicates that the link to that neighbor is currently unknown. Initially, the states of the links to the neighbors are unknown.
%           To change this state to the confirmed, an \it{RREP\_Ack} is requested whenever a \it{rrep} message is sent to determine if the node is
%           adjacent to the chosen next-hop.
            \item \it{Blacklisted}: indicates that the link to that neighbor is unidirectional. When a node has failed to receive the \it{RREP\_Ack} message in response to its \it{rreq} message to that neighbour, the  neighbor state is changed to blacklisted. Hence, it stops forwarding any message to it for an amount of time, \it{ResetTime}. After reaching the \it{ResetTime}, the neighbor's state will be set to \it{unknown}.
        \end{itemize}
    \item route table: contains information about discovered routes and their status: The following information is maintained for each route:
    \begin{itemize}
        \item\it{SeqNum}: destination sequence number
        \item\it{route\_state}: the state of the route to the destination which can have one of the following values:
        \begin{itemize}
            \item \emph{unconfirmed}: when the neighbor state of the next hop is unknown;
            \item \emph{active}: when the link to the next hop has been confirmed, and the route is currently used;
             \item \emph{idle}: when the link to the next hop has been confirmed, but it has not been used in the last \it{ACTIVE\_INTERVAL};
            \item \emph{invalid}: when the link to the next hop is broken, i.e., the neighbor state of the next hop is blacklisted.
        \end{itemize}
        \item \it{Metric}: indicates the cost or quality of the route, e.g., hop count, the number of hops to the destination
        \item \it{NextHop}: IP address of the next hop to the destination
        \item \it{Precursors} (optional feature): the list of the nodes interested in the route to the destination, i.e., upstream neighbors.
    \end{itemize}
    \item route message table, also known as \emph{RteMsg Table}: contains information about previously received route messages such as ${\it rreq}$ and ${\it rrep}$, so that we can determine whether the new received message is worth processing or redundant. Each entry of this table contains the following information:
    \begin{itemize}
        \item \it{MessageType}: which can be either ${\it rreq}$ or ${\it rrep}$
        \item \it{OrigAdd}: IP address of the originator
        \item \it{TargAdd}: IP address of the destination
        \item \it{OrigSeqNum}: sequence number of the originator
        \item \it{TargSeqNum}: sequence number of the destination
        \item \it{Metric}
    \end{itemize}
\end{itemize}
%todo::why we have abstract it

When one node, i.e., source, intends to send a package to another, i.e., destination, it looks up its
routing table for a valid route to that destination, i.e., a route of which the route state is not invalid. If there is no
such a route, it initiates a route discovery procedure by
broadcasting a $\it rreq$ message. The freshness of the requested route is indicated through the
sequence number of the destination that the source is aware of. Whenever a node initiates a route discovery, it
increases its own sequence number, with the aim to define the freshness of its route request. Every node upon receiving this
message checks its routing table for finding a route to the
requested destination. If there is such a path or the receiver is in
fact the destination, it informs the sender through unicasting a
$\it rrep$ message. However, an acknowledgment is requested whenever the neighbor state of the next hop is \emph{unconfirmed}. Otherwise, it re-broadcasts  the
$\it rreq$ message to examine if any of its neighbors has a valid
path. Meanwhile, a reverse forwarding path is constructed to the
source over which $rrep$ messages are going to be communicated later. In case a
node receives a $\it rrep$ message, if it is not the source, it
forwards the $\it rrep$ after updating its routing
table with the received route information.
%, otherwise it starts forwarding the
%data through the previously constructed route.
Whenever a node
fails to receive a requested ${\it RREP\_Ack}$, it uses a $\it rerr$ message to
inform all its neighbors intended to use the broken link to forward
their packets.

  %In addition, route discovery messages
%are identified by the combination of the sequence number and IP of
%the originator node.

In our model, each node is represented through a rebec (actor),
identified by an IP address, with a routing table and a sequence number ($\it sn$).
In addition, every node keeps track of the adjacency status to its
neighbors by means of a neighbor table, through the ${\it neigh\_state}$ array, where ${\it neigh\_state}[i]={\it
true}$ indicates that it is adjacent to the node with the IP address $i$, while
${\it false}$ indicates that its adjacency status is either
\emph{unknown} or \emph{blacklisted} (since timing issues are not taken into account, these two statuses are considered the same). %The node state variables are
%defined in the following. %For the sake of simplicity route message table is combined together with the routing table and is modeled as a set of arrays.
As the destinations of any two arbitrary rows of a routing table are always different, the routing table has at most $n$ rows, where $n$ is the number of nodes in the model. Therefore, the routing table is modeled by a set of arrays, namely, $\it dsn$, ${\it route\_state}$, $\it hops$, $\it nhops$, and $\it pres$, to represent the $\it SeqNum$, ${\it route\_state}$, $\it Metric$, $\it NextHop$, and $\it Precursors$ columns of the routing table, respectively.   The arrays $\it dsn$ and ${\it route\_state}$ are of size $n$, while the arrays $\it hops$, $\it nhops$, and $\it pers$ are of size $n\times n$. For instance, ${\it dsn}[i]$, keeps the sequence number of the destination with IP address $i$, while ${\it nhops}[i][j]$ contains the next hop of the $j$-th route to the destination with the IP address $i$.
%Each
%of the following information which collectively determines the freshness
%and length of each path to a destination:% is maintained in an array:
\begin{itemize}
\item$\it dsn$: destination sequence number
\item${\it route\_state}$: an integer that refers to the state of the route to the destination and can have one of the following values:
\begin{itemize}
    \item ${\it route\_state}[i]=0$: the route is \emph{unconfirmed}, there may be more than one route to the destination $i$ with different next hops and hop counts;
    \item ${\it route\_state}[i]=1$: the route is \emph{valid}, the link to the next hop has been confirmed, the route state in the protocol is either \textbf{active} or \textbf{idle}; since we abstract from the timing issues, these two states are depicted as one;
    \item ${\it route\_state}[i]=2$: the route is \emph{invalid}, the link to the next hop is broken;
\end{itemize}
\item $\it hops$: the number of hops to the destination for different routes
\item $\it nhop$: IP address of the next hop to the destination for different routes
\item $\it pres$: an array that indicates which of the nodes are interested in the routes to the destination, for example ${\it pres}[i][j]={\it true}$ indicates that the node with the IP address $j$ is interested in the routes to the node with the IP address $i$.
\end{itemize}
%Since we have considered a row for each destination in our routing table, to indicate whether the node has any route to each destination until now, each rebec has an auxiliary array called $\it exist$. For instance, ${\it exist}[i]={\it false}$ implies that the node has never had known any route to the node with the IP $i$.
Since we have considered a row for each destination in our routing
table, to indicate whether the node has any route to each
destination until now, we initially set ${\it dsn}[i]$ to $-1$
which implies that the node has never known any route to the
node with the IP address $i$. We refer to the all above mentioned arrays as
\emph{routing arrays}. Initially all integer cells of arrays are set
to $-1$ and all boolean cells are set to ${\it false}$. To model
expunging a route, its corresponding next hop and hop count entries
in the arrays $\it nhops$ and $\it hops$ are set to $-1$. Since we
have only considered one node as the destination and one node as the source, the
information in $\it rreq$ and $\it rrep$ messages has no conflict and
consequently the route message table can be abstracted away. In other words, the routing table information can be used to identify whether the new received message has been seen before or not, as the stored routes towards the source represent information about $\it rreq$s and the routes towards the destination represent $\it rrep$s. 
%For the sake of simplicity route message table is combined together with the routing table and is modeled as a set of arrays.

Note that $\it rreq$ and $\it rrep$, i.e., all route messages, carry route information to their source and destination, respectively. Therefore, a bidirectional path is constructed while these messages travel through the network. Whenever a node receives a route message, it processes incoming information to determine whether it offers any improvement to its known existing routes. Then, it updates its routing table accordingly in case of an improvement. The processes of evaluating and updating the routing table are explained in the following subsections.

\subsection{Evaluating Route Messages}\label{subsec::evalmsg}
Every received route message contains a route and consequently is evaluated to check for any improvement. Note that a $\it rreq$ message contains a route to its source while a $\it rrep$ message contains a route to its destination. Therefore, as the routes are identified by their destinations (denoted by $\it des$), in the former case, the destination of the route is the originator of the message (i.e., ${\it des}={\it oip}\_$), and in the latter, it is the destination of the message (i.e., ${\it des}={\it dip}\_$). The routing table must be evaluated if one of the following conditions is realized:
\begin{enumerate}
    \item no route to the destination has existed, i.e., ${\it dsn}[{\it des}]={\it -1}$
    \item there are some routes to the destination, but all their route states are \emph{unconfirmed}
    \item there is a valid or invalid route to the destination in the routing table and one of following conditions holds:
    \begin{itemize}
        \item the sequence number of the incoming route is greater than the existing one
        \item the sequence number of the incoming route is equal to the existing one, however the hop count of the incoming route is less than the existing one (the new route offers a shorter path and also is loop free)
    \end{itemize}
\end{enumerate}

\subsection{Updating the Routing Table}\label{subsec::updroutetable}
The routing table is updated as follows:
\begin{itemize}
\item if no route to the destination has existed, i.e., ${\it dsn}[{\it des}]={\it -1}$, the incoming route is added to the routing table.
\item if the route states of existing routes to the destination are \emph{unconfirmed}, the new route is added to the routing table. %So, the routing table may have multiple \emph{unconfirmed} routes.
%If an Unconfirmed route becomes valid in future, any remaining Unconfirmed routes which would not offer improvement will be expunged.
\item the incoming route has a different next hop from the existing one in the routing table, while the next hop's neighbor state of the incoming route is \emph{unknown} and the route state of the existing route is \emph{valid}. The new route should be added to the routing table since it may offer an improvement in the future and turn into \emph{confirmed}.
%If AdvRte.NextHop is not equal to Route.NextHop, and AdvRte.NextHop’s Neighbor.State is Unknown and Route.State is Active or Idle, the current route is valid but the advertised route may offer improvement, if it can be confirmed. Continue to Step 3 and create a new route table entry. It can replace the original route when Neighbor.State is set to Confirmed.
\item if the existing route state is \emph{invalid} and the neighbor state of the next hop of the incoming route is \emph{unknown}, the existing route should be updated with information of the received one.
%If AdvRte.NextHop’s Neighbor.State is Unknown and Route.State is Invalid, continue to Step 4 and update the existing route table entry.
\item if the next hop's neighbor state of the incoming route is \emph{confirmed}, the existing route is updated with new information and all other routes with the route state \emph{unconfirmed} are expunged from the routing table.
%If AdvRte.NextHop’s Neighbor.State is Confirmed, continue to Step 4 and update the existing route table entry
\end{itemize}

As described earlier, there are three types of route
discovery packets: $\it rreq$, $\it rrep$ and $rerr$. There is a
message server for handling each of these packet types:
\begin{itemize}
\item ${\it rec\_rreq}$ is responsible for processing a route discovery request message;
\item ${\it rec\_rrep}$ handles a reply request message;
\item ${\it rec\_rerr}$ updates the routing table in case an error occurs over a path and informs the interested nodes about the broken link.
\end{itemize}

There are also two message servers for receiving and sending a data
packet. All these message servers will be discussed thoroughly in
the following subsections.
%
%\begin{figure}
%\begin{center}
%%\begin{align*}
%\begin{lstlisting}[language=rebeca]
%int sn;                        //node_sequence_number int ip; int[]
%dsn;                   //destination sequence number int[]
%route_state; bool[] neigh_state; int[][] hops; int[][] nhop;
%bool[][] pre;       //all nodes that are interested in the route to
%dip are set to true bool[] exist; int[][] rec_bid; int[] store;
%//pointer: dip=>data for that dip
%\end{lstlisting}
%%\end{align*}
%\end{center}
%\vspace{-2mm}
%\caption{The state variables of a node in AODV protocol
%\label{code:AODVstateVars}}
%\end{figure}

\subsection{$\it rreq$ Message Server}\label{subsec::rreq}
This message server processes a received route
discovery request and reacts based on its routing table, shown in
\figurename{~\ref{code:rreq}}. The ${\it rreq}$ message has the
following parameters: $\it hops\_$ and $\it maxHop$ as the number of hops and the maximum number of hops, $\it dsn\_$ as the
destination sequence number, and ${\it oip\_}$, ${\it osn\_}$, ${\it dip}$, and ${\it sip\_}$
respectively refer to the IP address and sequence number of the originator, and the IP address of the destination, and the IP address of the sender. Whenever a node receives a route request, i.e., ${\it rec\_rreq}(hops\_, dip\_ ,dsn\_ ,oip\_ ,osn\_ ,sip\_,maxHop)$
message, it checks incoming information with the aim to improve the
existing route or introduce a new route to the destination, and then
updates its routing table accordingly (see also Sections \ref{subsec::evalmsg} and \ref{subsec::updroutetable}). During processing an $\it rreq$
message, a backward route, from the destination to the originator is
built by manipulating the routing arrays with the index $oip\_$. Similarly, while
processing an $\it rrep$ message, it constructs a forwarded route to the
destination by addressing the routing arrays with the index $dip\_$. Therefore, the procedure of evaluating the new route and updating the routing table is the
same for both $\it rreq$ and $\it rrep$ messages, except for different
indices $oip\_$ and $dip\_$, respectively.
\vspace{2mm}

\noindent{\bf{Updating the routing table:}}~\figurename{~\ref{code:UpdatingRoutingTable}} depicts this procedure which includes both evaluating the incoming route and updating the routing table (the code is the body of ${\it if}$-part in the line $7$ of \figurename{~\ref{code:rreq}}). If no route exists to the destination, the received information is used to update the routing table and generate discovery packets, lines (1-10). The route state is set based on the neighbor status of the sender: if its neighbor status is \emph{confirmed}, the route state is set to \emph{valid}, otherwise to \emph{unconfirmed}. The next hop is set to the sender of the message, i.e.,\ $nhop[oip\_][0]=sip\_$. If a route exists to the destination (i.e., ${\it oip}\_$ ), one of the following conditions happens:
\begin{itemize}
    \item the route state is \emph{unconfirmed}, lines (11-36): it either updates the routing table if there is a route with a next hop equal to the sender, or adds the incoming route to the first empty cells of $nhop$ and $hops$ arrays.
    If the neighbor status of the sender is \emph{confirmed}, then all other routes with the same destination are expunged while the route state is set to \emph{valid}, lines (21-30).
    \item the route state is \emph{invalid} or it is valid, but the neighbor status of the sender is \emph{confirmed}, lines (38-48): if the incoming message contains a greater sequence number, or an equal sequence number with a lower hop count, then it updates the current route while a new discovery message is generated.
    \item the route state is \emph{valid} and the neighbor status of the sender is \emph{unknown}, lines (50-66): the incoming route is added to the routing table and a new discovery message is generated if it provides a fresher or shorter path.
\end{itemize}

\begin{figure}
    \begin{center}
        \begin{lstlisting}[language=rebeca]
if(dsn[oip_]==-1) {
    dsn[oip_]=osn_;
    if(neigh_state[sip_]==true)
       { route_state[oip_]=1; }
    else
       { route_state[oip_]=0; }
    hops[oip_][0]=hops_;
    nhop[oip_][0]=sip_;
    gen_msg = true;
} else {
    if(route_state[oip_]==0) {
        dsn[oip_]=osn_;
        route_num = 0;
        for(int i=0;i<4;i++) 
        {
            if(nhop[oip_][i]==-1 || nhop[oip_][i]==sip_) {
                route_num = i;
                break;
            }
        }
        if(neigh_state[sip_]==true) {
            route_state[oip_]=1;
            for(int i=0;i<4;i++) 
            {
                hops[oip_][i]=-1;
                nhop[oip_][i]=-1;
            }
            hops[oip_][0]=hops_;
            nhop[oip_][0]=sip_;
        }
        else {
            route_state[oip_]=0;
            hops[oip_][route_num]=hops_;
            nhop[oip_][route_num]=sip_;
        }
    }
    else {
        if(route_state[oip_]==2 || neigh_state[sip_]==true) {
            /* update the existing route */
            if((dsn[oip_]==osn_ && hops[oip_][0]>hops_) || dsn[oip_]<osn_ ) {
                dsn[oip_]=osn_;
                if(neigh_state[sip_]==true)  route_state[oip_]=1; 
                   else route_state[oip_]=0; 
                hops[oip_][0]=hops_;
                nhop[oip_][0]=sip_;
                gen_msg = true;
            }
        }
        else {
			route_num = 0;
			for(int i=0;i<4;i++)
			{
				if(nhop[oip_][i]==-1 || nhop[oip_][i]==sip_)
				{
					route_num = i;
					break;
				}
			}
			if((dsn[oip_]==osn_ && hops[oip_][0]>hops_) || dsn[oip_]<osn_ )
			{
				 dsn[oip_]=osn_;
				 hops[oip_][route_num]=hops_;
				 nhop[oip_][route_num]=sip_;
				 gen_msg = true;
			}
	    }
	}
}
        \end{lstlisting}
    \end{center}
    \vspace{-2mm}
    \caption{Updating the routing table
        \label{code:UpdatingRoutingTable}}
\end{figure}

In these cases, if a new discovery message should be generated (when the node has no route as fresh as the route request), the auxiliary boolean variable $gen\_msg$ is set to $\it true$. In \figurename{~\ref{code:rreq}}, after updating the routing table, if a new message should be
generated, indicated by ${\it if}~(gen\_msg=true)$, it rebroadcasts the
$\it rreq$ message with the increased hop count if the node is
not the destination, lines (51-54). Otherwise, it increases its
sequence number and replies to the next hop(s) toward the originator
of the route request, $oip\_$, based on its routing table. Before
unicasting $\it rrep$ messages, next hops toward the
destination, $dip\_$, and the sender are set as interested nodes to
the route toward the originator, $oip\_$, lines (17-22). It unicasts
each $\it rrep$ message to its next hops one by one until it gets
an ack from one, lines (23-43); ack reception is modeled implicitly
through successful delivery of unicast, i.e., the $succ$ part. If it
receives an ack, it updates the route state to \emph{valid} and the
neighbor status of the next hop to \emph{confirmed} and stops
unicasting $\it rrep$ messages. If it doesn't receive an ${\it RREP\_Ack}$ message from the next hop when the route state is \emph{valid}, it initiates the error recovery procedure.
\vspace{2mm}

\noindent{\bf{Error Recovery Procedure:}} The code for this
procedure is illustrated in
\figurename{~\ref{code:ErrorRecoveryProcedure}} (its code is the body of ${\it if}$-part in line $46$ of \figurename{~\ref{code:rreq}}). As explained earlier, this procedure is initiated when a node doesn't receive an ${\it RREP\_Ack}$ message from the next hop of the route with state \emph{valid}. Then, it updates its route state to \emph{invalid} and adds the sequence number of the originator to the array of invalidated sequence numbers, denoted by ${\it dip\_sqn}$. Furthermore, it
adds all the interested nodes in the current route to the list of
affected neighbors, denoted by ${\it affected\_neighbours}$, lines (3-7). It invalidates other valid routes
that use the same broken next hop as their next hops, adds their
sequence numbers to the invalidated array and sets the nodes
interested in those routes as affected neighbors, lines (8-24).
Finally, it multicasts an $\it rerr$ message which contains the
destination IP address, the node IP address, and the invalidated sequence numbers to
the affected neighbors, line 25.

\begin{figure}
    \begin{center}
        \begin{lstlisting}[language=rebeca]
route_state[oip_]=2;
dip_sqn[oip_]=dsn[oip_];
for(int k=0;k<4;k++)
{
    if(pre[oip_][k]==true)
        { affected_neighbours[k]=true; }
}
for(int j=0;j<4;j++)
{
    for(int r=0;r<4;r++)
    {
        if(nhop[oip_][r]!=-1 &&‌ nhop[j][0]==nhop[oip_][r])
        {
            route_state [j]= 2;
            dip_sqn[j]=dsn[j];
            for(int k=0;k<4;k++)
            {
                if(pre[j][k]==true)
                    { affected_neighbours[k]=true; }
            }
            break;
        }
    }
}
multicast(affected_neighbours,rec_rerr(dip_,ip,dip_sqn));
        \end{lstlisting}
    \end{center}
    \vspace{-2mm}
    \caption{The error recovery procedure
        \label{code:ErrorRecoveryProcedure}}
\end{figure}

\begin{figure}
\begin{center}
\begin{lstlisting}[language=rebeca]
 msgsrv rec_rreq(int hops_,int dip_ ,int dsn_ ,int oip_ ,int osn_ ,int sip_,int maxHop)
{
    int[]  dip_sqn=new int[4];
    int route_num;
    boolean[] affected_neighbours=new boolean[4];
    boolean gen_msg = false;
    if(ip!=oip_)
    {
        //evaluate and update the routing table
    }
    if(gen_msg==true)
    {
        if(ip==dip_)
        {
            boolean su = false;
            pre[dip_][sip_]=true;
            for(int i=0;i<4;i++)
            {
                int nh = nhop[dip_][i];
                if(nh!=-1)
                    { pre[oip_][nh]=true; }
            }
            for(int i=0;i<4;i++)
            {
                if(nhop[oip_][i]!=-1)
                {
                    int n_hop = nhop[oip_][i];
                    sn        = sn+1;
                    /* unicast a RREP towards oip of the RREQ */
                    unicast(n_hop,rec_rrep(0 , dip_ , sn , oip_ , self))
                    succ:
                    {
                        route_state[oip_]=1;
                        neigh_state[n_hop]=true;
                        su = true;
                        break;
                    }
                    unsucc:
                    {
                        neigh_state[n_hop]=false;
                    }
                }
            }
            if(su==false && route_state[oip_]==1)
            {
                /* error recovery procedure */
            }
        }
        else {
            hops_ = hops_+1;
            if(hops_<maxHop)
                { rec_rreq(hops_,dip_,dsn_,oip_,osn_,self,maxHop); }
        }
    }
}
\end{lstlisting}
\end{center}
\vspace{-2mm} \caption{The rreq message server \label{code:rreq}}
\end{figure}

\subsection{$\it rrep$ Message Server}
This message server, shown in \figurename{~\ref{code:rrep}}, processes the received reply messages and also
constructs the route forward to the destination. At first, it
updates the routing table and decides whether the message is worth
processing, as previously mentioned for $\it rreq$ messages, and constructs the route, but this time to the destination (its code is similar to the one in \figurename{~\ref{code:UpdatingRoutingTable}} except that $\it dip\_$ is used instead of $\it oip\_$, and is place at line $6$ of \figurename{~\ref{code:rrep}}). This message is sent backwards till it
reaches the source through the reversed path constructed while
broadcasting the $\it rreq$ messages. When it reaches the source,
it can start forwarding data to the destination. In case the node is
not the originator of the route discovery message, it updates the
array of interested nodes, lines (17-23). Then, it unicasts the message to the next hop(s), on the reverse path to the originator, lines (24-42). Based on the AODVv2 protocol, if connectivity to the next hop on the route to the originator is not confirmed yet, the node must request a Route Reply Acknowledgment ({\it RREP\_Ack}) from the intended next hop router. If a {\it RREP\_Ack} is received, then the neighbor status of the next hop and route state must be updated to \emph{confirmed} and \emph{valid}, respectively, lines (30-36), otherwise the neighbor status of the next hop remains \emph{unknown}, lines (37-40). This procedure is modeled through \textit{conditional unicast} which enables the model to react based on the delivery status of the unicast message so that $succ$ models the part where the {\it RREP\_ACK} is received while $unsucc$ models the part where it fails to receive an acknowledgment from the next hop. In case the unicast is unsuccessful and the route state is valid, the error recovery procedure will be followed, lines (43-46).%If all the unicasts fail, the error recovery procedure will be followed.

\begin{figure}
\begin{center}
\begin{lstlisting}[language=rebeca]
msgsrv rec_rrep(int hops_ ,int dip_ ,int dsn_ ,int oip_ ,int sip_) {
    int[]  dip_sqn=new int[4];
    boolean[] affected_neighbours=new boolean[4];
    boolean gen_msg = false;
    int n_hop,route_num;
    /* evaluate and update the routing table */
    if(gen_msg==true)
    {
        if(ip==oip_ )
        {
            /* this node is the originator of the corresponding RREQ */
            /* a data packet may now be sent */
        }
        else {
            hops_   = hops_+1;
            boolean su = false;
            pre[oip_][sip_]=true;
            for(int i=0;i<4;i++)
            {
                n_hop = nhop[oip_][i];
                if(n_hop!=-1)
                    { pre[oip_][n_hop]=true; }
            }
            for(int i=0;i<4;i++)
            {
                if(nhop[oip_][i]!=-1)
                {
                    n_hop = nhop[oip_][i];
                    unicast(n_hop,rec_rrep(hops_,dip_,dsn_,oip_,self))
                    succ:
                    {
                        route_state[oip_]=1;
                        neigh_state[n_hop]=true;
                        su = true;
                        break;
                    }
                    unsucc:
                    {
                        neigh_state[n_hop]=false;
                    }
                }
            }
            if(su==false && route_state[oip_]==1)
            {
                /* error recovery procedure */
            }
        }
    }
}
\end{lstlisting}
\end{center}
\vspace{-2mm} \caption{The rrep message server \label{code:rrep}}
\end{figure}

\subsection{$\it rerr$ Message Server}
This message server, shown in \figurename{~\ref{code:rerr}}, processes the received error messages and
informs those nodes that depend on the broken link. When a node
receives an $\it rerr$ message, it must invalidate those routes
using the broken link as their next hops and sends the $\it rerr$
message to those nodes interested in the invalidated routes. This
message has only two parameters: $sip\_$ which indicates the IP address of the sender, and $rip\_rsn$, which contains the sequence number of those destinations which have become unaccessible from the $sip\_$.

For all the \emph{valid} routes to the different destinations, it examines whether the next hop of the route to the destination is equal to $sip\_$ and the sequence number of the route is smaller then the received sequence number, line 10. In case the above conditions are satisfied, the route is invalidated, lines (11-19), and an $rerr$ message is sent to the affected nodes, line 21.

\begin{figure}
\begin{center}
\begin{lstlisting}[language=rebeca]
msgsrv rec_rerr(int source_, int sip_ ,int[] rip_rsn) {
    int[]  dip_sqn=new int[4];
    boolean[] affected_neighbours=new boolean[4];
    if(ip!=source_)
    {
        //regenerate rrer for invalidated routes
        for(int i=0;i<4;i++)
        {
            int rsn=rip_rsn[i];
            if(route_state[i]==1 && nhop[i][0]==sip_ && dsn[i]<rsn && rsn!=0)
            {
                route_state [i]= 2;
                dip_sqn[i]=dsn[i];
                for(int j=0;j<4;j++)
                {
                    if(pre[i][j]==true)
                        { affected_neighbours[j]=true; }
                }
            }
        }
        multicast(affected_neighbours,rec_rerr(source_,self,dip_sqn));
    }
}
\end{lstlisting}
\end{center}
\vspace{-2mm} \caption{The rerr message server \label{code:rerr}}
\end{figure}

\subsection{$\it newpkt$ Message Server}
Whenever a node intends to send a data packet, it creates a
$rec\_newpkt$ which has only two parameters, $data$
and $dip\_$. The code for this message
server is shown in \figurename{~\ref{code:recNewpkt}}. If it is the destination of the message, it
delivers the message to itself, lines (4-7). Otherwise, if it has a valid route
to the destination, it sends data using that
route, lines (11-15). If it has no valid route, it increases its own sequence number
and broadcasts a route request message, lines (16-25).
In addition, if a route to the destination is not found within ${\it RREQ}\_{\it WAIT}\_{\it TIME}$, the node retries to send a new $\it rreq$ message after increasing its own sequence number. Since we abstracted away from time, we model this procedure through the ${\it resend\_rreq}$ message server which attempts to resend an $\it rreq$ message while the node sequence number is smaller than $3$ (to make the state space finite).
%RREQ_Gen awaits reception of a Route Reply message (RREP) containing a route toward TargAddr. An RREQ from TargAddr would also fulfil the request, if adjacency to the next hop is already confirmed. If a route to TargAddr is not learned within RREQ_WAIT_TIME, RREQ_Gen MAY retry the route discovery. To reduce congestion in a network, repeated attempts at route discovery for a particular target address SHOULD utilize a binary exponential backoff: for each additional attempt, the waiting time for receipt of the RREP is multiplied by 2. If the requested route is not learned within the wait period, another RREQ MAY be sent, up to a total of DISCOVERY_ATTEMPTS_MAX. This is the same technique used in AODV [RFC3561].
\begin{figure}
\begin{center}
\begin{lstlisting}[language=rebeca]
msgsrv rec_newpkt(int data ,int dip_) {
    int[]  dip_sqn=new int[4];
    boolean[] affected_neighbours=new boolean[4];
    if(ip==dip_ )
     {
        /* the DATA packet is intended for this node */
    }
    else{
        /* the DATA packet is not intended for this node */
        store[dip_]=data;
        if(route_state[dip_]==1)
        {
            /* valid route to dip*/
            /* forward packet */
        }
        else{
            /* no valid route to dip*/
            /* send a new rout discovery request*/
            if(sn<3)
            {
                sn++;
                unicast(self,resend_rreq(dip_));
                rec_rreq(0,dip_,dsn[dip_],self,sn,self,4);
            }
        }
    }
}
msgsrv resend_rreq(int dip_)
{
    if(sn<3)
    {
        sn++;
        unicast(self,resend_rreq(dip_));
        rec_rreq(0,dip_,dsn[dip_],self,sn,self,4);
    }
}
\end{lstlisting}
\end{center}
\vspace{-2mm} \caption{The rec\_newpkt message server
\label{code:recNewpkt}}
\end{figure}

%% file: Evaluation.tex
\section{Evaluation} \label{sec::eval}
In this section, we will review the results obtained from efficiently
constructing the state spaces for the two introduced wRebeca models, the
flooding and AODV protocol. Also, we briefly introduce our tool and
its capabilities. Then, the loop freedom invariant is defined and one possible loop scenario is demonstrated. Finally, two properties that must hold for the AODV protocol are expressed that can be checked with regard to the AODV model.

\subsection{State-Space Generation}
\noindent{\bf{Static Network .}} Consider a network with a static topology, in other words the network constraint is defined so that it leads to only one valid topology. We illustrate the applicability of our counting abstraction technique on the flooding routing protocol. In contrast to the intermediate nodes on a path (the ones except the source and destination), the two source and destination nodes cannot be aggregated (due to their local states). However, in the case of the AODV protocol, no two nodes can be counted together due to the unique variables of IP and routing table of each node. As the number of intermediate nodes with the same neighbors increases, the more reduction takes place. We have precisely chosen four fully connected network topologies to show the power of our reduction technique when the intermediate nodes increase from one to four.

\tablename{~\ref{tab:compareStateSpaceFlooding}} illustrates the
number of states when running the flooding protocol on different networks
with different topologies before and after applying counter
abstraction reduction.  In the first, second, third, and fourth topology, there are three nodes with one intermediate, four nodes with two, five nodes with three, and six with four intermediates, respectively.
%,  nodes 3 and 1 are both connected to 0 and 2 while node 0 wants to send a packet to node 2. Therefore node 3 and 1 are intermediate nodes and have the same role in the flooding protocol, they can pass a sent packet from node 0 to node 2.
By applying counter abstraction reduction, the intermediate nodes are collapsed together as they have the same role in the protocol. However, the effectiveness of this technique depends on the network topology and the modeled protocol. %Nevertheless, if no nodes have similar neighbor, as a result there would be no reduction, as it is expected.

\begin{table}
    \centering
        \caption{Comparing the size of state spaces with/without applying counter abstraction reduction}

    \begin{tabular*}{0.8\textwidth}{@{\extracolsep{\fill }}  | c  r  r  r  r |  }
    \hline
          No. of   & No. of states&  No. of
         states & No. of transitions & No. of transitions
    \\
           intermidate  nodes &  before reduction & after reduction & before  reduction &‌ after reduction \\
         \hline    
      1 &    24 & 24 & 36 & 36 \\
      2 &   226 & 133 & 574 & 276 \\
      3 &   3,689 & 912 & 13,197 & 2,441 \\
      4 &   71,263 & 6,649 & 321,419 & 21,466 \\
        \hline
    \end{tabular*}
    \label{tab:compareStateSpaceFlooding}
\end{table}

\vspace*{2mm}

\noindent{\bf{Dynamic network.}} At these networks, topology is constantly changing, in other words there are more than one possible topology. The resulting state spaces after and
before eliminating $\tau$-transitions are compared for the two case
studies while the topology is constantly changing for a networks of
$4$ and $5$ nodes, as shown in
\tablename{~\ref{tab:compareStateSpaceWithAndWithoutTau}}.
\tablename{~\ref{tab:constraints}} depicts the constraints used to
generate the state spaces and the number of topologies that each
constraint results in. Constraints are chosen randomly here, just to show the effectiveness of our reduction technique. To this aim, we have randomly removed a (fixed) link from the network constraints.  Nevertheless, constraints can be chosen wisely to limit the network topologies to those which are prone to lead to an erroneous situation, i.e., violation of a correctness property like loop freedom. However, it is also possible to check the model against all possible topologies by not defining any constraint. In other words, a modeler at first can focus on some suspicious network topologies and after resolving the raised issues it check the model for all possible topologies.  There are also some networks which have certain constraints about how the topology can change, e.g., node $1$ can never get into the communication range of node $2$. These restrictions on topology changes can be reflected through constraints too.
The sizes of state spaces are compared under
different network constraints resulting in different number of valid
topologies. Eliminating $\tau$-transitions and topology information
manifestly reduces the number of states and transitions even when
all possible topologies are not restricted. Therefore, it makes
MANET protocol verification possible in an efficient manner. Note that in case the size of the network was increased from four to five, we couldn't generate its state space without applying reduction due to the memory limitation on a computer with $8$GB RAM.

\begin{table}
    \centering
        \caption{Comparing the size of state spaces with/without applying $\tau$-transition elimination reduction}
    \begin{tabular*}{0.95\textwidth}{@{\extracolsep{\fill }} | c  c  c  r  r  r  r  |   }
    \hline
         &  No. of & No. of valid & No. of states    &  No. of transitions     & No. of states & No. of transitions
        \\
         &  nodes & topologies & before reduction  &  before reduction   & after reduction &  after reduction
        \\
         \hline
         flooding    
         & 4 & 4 & 2,119 & 11,724 & 541 & 1,652 \\
         protocol & 4  &  8 & 4,431 & 42,224 & 567  & 1,744 \\
         & 4& 16 & 10,255  & 179,936 & 655  & 2,192\\
         & 4& 32 & 22,255 & 747,200 & 710 & 2,765\\
         & 4        & 64  & 44,495 & 2,917,728 & 710 & 3,145\\
         \hline
         AODV     
         &   4 & 4 & 3,007 & 16,380 & 763  & 1,969 \\
         protocol  & 4  & 8 & 12,327 & 113,480 & 1,554  & 3,804 \\
         &  4  & 16 & 35,695 & 610,816 & 2,245 & 5,549 \\    
         &  4  & 32 & 93,679 & 3,097,792 & 2,942 & 7,596 \\  
         &  4    & 64  & 258,447  & 16,797,536 & 4,053 & 10,629 \\
         &     5 & 16 & $>$655,441 & $>$11,276,879 & 165,959 &  598,342 \\
         \hline
    \end{tabular*}
    \label{tab:compareStateSpaceWithAndWithoutTau}
\end{table}

\begin{table}
    \centering
    \caption{Applied network constraints}
    \begin{tabular*}{0.83\textwidth}{@{\extracolsep{\fill }} |  c  c   l   |}
        \hline
       No. of  &   No. of valid & constraint
        \\
      nodes &  topologies & \\
        \hline
     4 &  4 &‌‌ $and (and (con (node0, node1), con (node0, node3)),$ \\
     &‌ &
    \hspace{1cm} $and (con (node2, node3), con (node1, node3)))$ \\
      4 &  8 & $and (and (con (node0, node1),$ \\
      & & \hspace{1cm}$con (node0, node3)), con(node2, node3))$\\
      4 & 16 & $and (con (node0, node1), con(node2, node3))$\\
        4 & 32 & $con (node0, node1)$\\
     5 &  16 & $and(and(con(node0,node1),and(con(node0,node3),con(node4,node1))),$ \\
      & &\hspace{1cm} $and(con(node2,node3),and(con(node1,node3),con(node2,node4))))$  \\
        \hline
    \end{tabular*}
    \label{tab:constraints}
\end{table}

\subsection{Tool Support}
\begin{figure}
   \centering
   \includegraphics[width=0.85\textwidth]{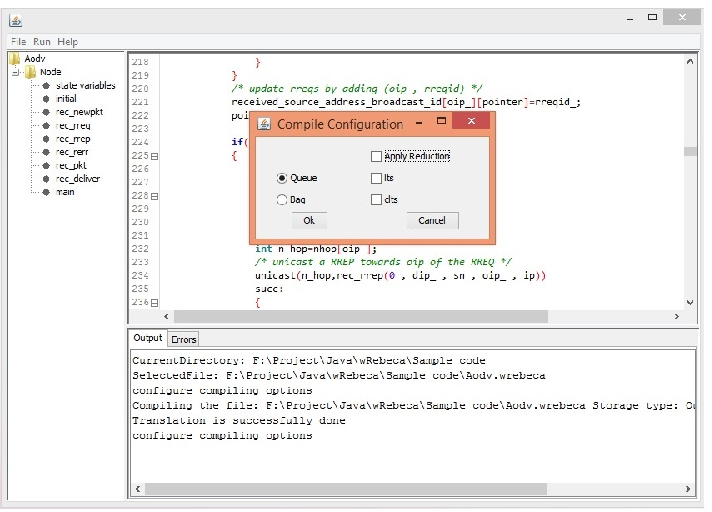}
   \caption{A screen-shot of the wRebca tool with the \textit{compilation info} window to configure the state-space generator}
   \label{fig:program}
   \end{figure}

The presented modeling language is supported by a tool (available at
\cite{BehnazJava}), providing a number of options to generate the
state space. A screen-shot of this tool is given in
\figurename{~\ref{fig:program}}. This tool supports
both bRebeca and wRebeca models characterized by different file
types. After opening a model, the tool extracts the information of the reactive classes, such as the state variables and message servers, and
also the main part including the rebec declarations and the network constraint.
Then it generates several classes in the Java language based on the
obtained information and compiles them together with some abstract
and base classes (common in all models), for example \textit{global
state} and \textit{topology}, to build an engine that constructs the
model state space upon its execution. Before compiling, a user can
decide about rebecs message processing method, in a FIFO manner
(queue) or in an arbitrary way (bag), and if the reduction should be
applied. To take advantage of all hardware capabilities, we have
implemented our state-space generation algorithm in a multi-threaded
way to leverage the power of multi-core CPUs.

During state-space generation, information about the state variables
and transitions are stored as an LTS in the \textsf{Aldebaran} format\footnote{\url{http://cadp.inria.fr/man/aldebaran.html}}. This LTS can be evaluated by tools such as the {\it mCRL2}
toolset \cite{mcrlTool}. For example, one can express desired
properties in $\mu$-calculus \cite{regular-mucalculus} and verify
them. Also, as explained in Section \ref{sec::reduction}, labels are
extended with network constraints as defined in \cite{FatemehTCS} so
that the reduced LTS can be model checked with respect to underlying
topology \cite{ghassemi2013model}.

\subsection{Model Checking of the AODV Protocol Properties\label{subsec::properties}}
There are different ways to check a given property on a wRebeca
model. Invariant properties can be evaluated while generating the
state space by checking each reached global state against defined
invariants. Furthermore, the resulting state space can be model checked
by tools supporting Aldebaran format such as mCRL2 and CLTS model
checker.

\subsubsection{Checking the Loop Freedom Invariant}
Loop freedom is one of the well-known property which must hold
for all routing protocols such as the AODV protocol. For example,
consider the routes to a destination $x$ in the routing tables of
all nodes, where $node_0$ has a route to $x$ with the next hop
$node_1$, $node_1$ has a route to $x$ with the next hop $node_2$,
and $node_2$ has a route to $x$ with the next hop $node_0$. The
given example constructs a loop which consists of the three nodes,
$node_0$, $node_1$, and $node_2$. A state is considered \emph{loop
free} if the collective routing table entries of all nodes for each
pair of a source and destination do not form a loop.  As it was
mentioned earlier in AODV-v2-11, each route may have more than one
next hop when the adjacency states of the next hops are
\emph{unconfirmed}. Therefore, while the loop freedom of a state is
checked, one must take into account all next hops stored for each
route. Then, for each next hop check whether it leads to a loop or
not. A routing protocol deployed on a network is called loop free 
if all of its reachable states are loop free. In other words, loop freedom
property of a protocol is an invariant (which can be easily
specified by the ACTL-X fragment of $\mu$-calculus, and hence, is
preserved by the reduced semantic model). However, We have extended our state-space generator engine  to check invariants (specified by functions) over each newly generated global state on the fly by calling the functions provided by a user, i.e., the invariants. To this aim, we have specified the loop freedom invariant by a recursive function, to inspect for a given global state whether the next hops in the routing table entries of nodes collectively lead to a loop-formation scenario, as shown in \figurename{~\ref{alg:loopFreedom}}.
Therefore, whenever the state-space generator reaches to a new state, before proceeding any further, it checks whether any loop is formed on the forward/backward routes between the source and destination, by calling ${\it loop\_freedom}(4,1,\mbox{new Set}\langle {\it int}\rangle (1))$ and ${\it loop\_freedom}(1,4,\mbox{new Set}\langle {\it  int}\rangle (4)) $, as ${\it node}_4$ and ${\it node}_1$ are the destination and source respectively. If the loop freedom condition is violated, the ${\it loop\_freedom}$ function returns ${\it false}$, and the state generator engine stops while it returns the path which has led to the global state under consideration as a counter example. The function ${\it loop\_freedom}$ has three parameters: $\it des$ refers to
 the destination of the route, $\it cur$ refers to the IP address of the current node
 which is going to be processed and $\it visited$ is the list of IP addresses of
 those nodes which have been processed.

\begin{figure}
\begin{center}
%\begin{align*}
\begin{lstlisting}[language=java]
bool loop_freedom(des:int, cur:int, visited:Set<int>){
    for(int i=0; i<n; i++)
        if ((state.node(cur).nhops[des][i]!=-1) && (!visited.contains(state.node(cur).nhops[des][i]))
                    && loop_freedom(des,i,visited.add(i)))
            return true;
        else
            return false;
}
\end{lstlisting}
\end{center}
\caption{Checking loop freedom property on a global state: we have used a dot notation to access the array ${\it nhops}$
of the rebec with the identifier $i$, i.e., ${\it state}.{\it node}(i)$, where $\it state$ is the newly generated global state} \label{alg:loopFreedom}

\end{figure}

Although keeping more than one next hop for each route may increase
the route availability, it compromises its validity by violating the
loop freedom invariant in a network of at least four nodes with a
dynamic topology. Consider the network topology shown in
\figurename{~\ref{Fig::Net}}. The following scenario explains steps
that lead to the invariant violation.
\begin{enumerate}
\item
$node_2$ initiates a route discovery procedure for destination $node_3$ by broadcasting a $rreq$ message.
\item
$node_1$ and $node_4$ upon receiving the $rreq$ message, add a route
to their routing tables towards $node_2$ and store $node_2$ as their
next hop. Since it is the first time that these nodes have received
a message from $node_2$, the neighbor state of $node_2$ is set to
\emph{unconfirmed}. Therefore, the route state is
\emph{unconfirmed}.
\item
As $node_1$ and $node_4$ are not the intended destination of the
route request, they rebroadcast the $rreq$ message.
\item
$node_1$ receives the $rreq$ message sent by $node_4$ and since the
route to $node_2$ is \emph{unconfirmed} it adds $node_4$ as a new next
hop to $node_2$.
\item
$node_4$ also adds $node_1$ as the new next hop towards $node_2$
after processing the $rreq$ sent by $node_1$. At this point a loop
is formed between $node_1$ and $node_4$.
\item
$node_3$ receives the $rreq$ message sent by $node_1$ and since it
is the destination, it sends a $rrep$ message towards $node_1$.
\item $node_2$ moves out of the communication ranges of $node_1$ and $node_4$.
\item
$node_1$ receives  the $rrep$ message sent by $node_3$ and as the
route state towards $node_2$ is \emph{unconfirmed} it unicasts the
$rrep$ message one by one to the existing next hops, $node_2$ and $node_4$, till it gets an ack. Since $node_2$ has moved out of the communication ranges of $node_1$, no ack is received from $node_2$ and $node_2$ gets removed from the routing table as the next hop to $node_2$. Then, another $rrep$ is sent to $node_4$. Since $node_4$ is adjacent to $node_1$, it receives the message and
then sends an ack to $node_1$. Therefore, $node_1$ sets the neighbor
state of $node_4$ to \emph{confirmed} and subsequently the route
state towards $node_2$ to \emph{valid}. 
\item
$node_4$ by receiving the $rrep$ message from $node_1$ unicasts it
to its next hops $node_1$ and $node_2$ similar to
$node_1$. Since it fails to receive an ack from $node_2$ and receives one from $node_1$, it updates its routing table by validating $node_1$ as its
next hop to $node_2$.
\end{enumerate}

We have found the scenario in the wRebeca model with the network constraint resulting four topologies as indicated in Table \ref{tab:constraints}. However, this scenario was also found for all the network constraints described in the table. Furthermore, we can generalize the scenario to all networks with the same connectivity when the communications occur, and the same mobility scenario.

\subsubsection{Checking the Properties by mCRL2}
Sequence numbers are used frequently by the AODV protocol to evaluate
the freshness of routes. Therefore, it is important that each node's
sequence number increases monotonically. To this end, we manually
configured the state generator to add two self-loops to each state
with the label ${\it src\_sn}(x)$  to
monitor the sequence number of the source node, where $x$ is $\it sn$ of the source node, and the label ${\it info\_i\_dsn}(y,z)$ to trace the destination sequence number of routes to the source and destination for each node $i$ (i.e., the backward and forward routes to the destination of our model), where $y$ and $z$ are ${\it dsn}[{\it src}]$ and ${\it dsn}[{\it dst}]$ of node $i$, respectively. These properties are
expressed through the ACTL-X fragment of $\mu$-calculus as shown in \figurename{~\ref{prop:sn}}. The first
formula asserts a monotonic increase of
the source sequence number. The second formula assures the destination
sequence numbers stored in the routing table of $node_i$ are
increased monotonically, and must hold for the 
nodes in the model.

\begin{figure}
    \begin{center}\[
    \begin{array}{l}
    \forall x,y:\mathbb{N}((x>0 \wedge x<4  \wedge  y>0 \wedge y<x) \Rightarrow [{\it src\_sn(x)}.{\it true}^*.{\it src\_sn(y)}]{\it
        false})\vspace*{4mm}\\
\forall x,y,m,n:\mathbb{N}(x\geq0 \wedge x<4 \wedge y\geq0 \wedge y<4 \wedge \\
\hspace*{2.5cm}m\geq0 \wedge m<4 \wedge n\geq0 \wedge n<4 \wedge  \\
\hspace*{2.5cm}(m<x \vee n<y))\Rightarrow[{\it true}^*.{\it info\_i\_dsn}(x,y).{\it true}^*.{\it info\_i\_dsn}(m,n)]{\it
    false})
    \end{array}\]
    \end{center}
    \vspace{-2mm}
    \caption{$\mu$-calculus properties verified by mCRL2}
        \label{prop:sn}
\end{figure}

\subsubsection{Checking Packet Delivery Property by the CLTS Model Checker}
The CLTS model checker can be used to express and verify interesting
properties of MANET protocols dependent to the underlying topology specified in Constrained Action Computation Tree Logic (CACTL)
\cite{ghassemi2013model}, an extension of Action CTL \cite{ACTL}. The path quantifier \emph{All} in CACTL is parametrized by a multi-hop constrain over the topology, which specifies the pre-condition required for
paths of a state to be inspected. Therefore, a state satisfies $\A{\mu}\varphi$ if its paths over which the multi-hop constraint $\mu$ holds, also satisfy $\varphi$. It also contains the two temporal operators \emph{until} and \emph{weak until} to specify the path formulae $\phi\U{\chi}{\chi'}\phi'$ and $\phi\W{\chi}{\chi'}\phi'$ to denote a path over which states satisfying $\phi$ are met by actions of $\chi$ until a state satisfying $\phi'$ is met by actions of $\chi'$ (in case of weak until, the state satisfying $\phi'$ can never be met).

The important property of
\textit{packet delivery} in routing or information dissemination
protocols in the context of MANETs becomes: if there exists an
end-to-end route (multi-hop communication path) between two nodes
$A$ and $C$ \emph{for a sufficiently long period of time}, then
packets sent by $A$ will eventually be received by $C$
\cite{fehnker2013process}.  To specify such the property, inspired
from \cite{fehnker2013process} we revised our specification to
include data packet handling (to forward the packet to its next hop towards the destination) in addition to the route discovery packets
and their corresponding handlers. Therefore, whenever a node,
source, discovers a route to an intended destination, it starts
forwarding its data packet through the next hop specified in its routing
table. The data packet is forwarded by intermediate nodes to their
next hops. When the data packet reaches the
intended destination, it delivers the data to itself by unicasting the
$\it deliver$ message to itself. In case an intermediate node fails
to forward the message, the error recovery procedure is followed as
explained in Section \ref{sec::AODV}. Consequently, using the
following formula, we can verify packet delivery
property:
{\[\A{\it true}({\it true}\W{\neg{\it
rec\_newpkt}(0,4)}{{\it rec\_newpkt}(0,4)}\A{{\it n}_1\pconn {\it n}_4\wedge {\it n}_4\pconn
{\it n}_1}({\it true}\U{\tau}{{\it deliver}()}{\it true}))
\]}It expresses that as long as there is a stable multi-hop path from $n_1$ to $n_4$ and vice versa (specified by ${\it n}_1\pconn {\it n}_4\wedge {\it n}_4\pconn
{\it n}_1$),
any ${\it rec\_newpkt}(0,4)$ message is proceeded by a ${\it delivery}()$ message
after passing $\tau$-transitions which abstract away from other
message communications. By model checking the resulting CLTS of the AODVv2
model, we found a scenario in which the property does not hold. We explain this scenario in a network of three nodes $N_1$, $N_2$ and $N_3$, where node
$N_3$ is always connected to the nodes $N_1$ and $N_2$, while the connection
between the nodes $N_1$ and $N_2$ is transient. Therefore, the mobility of nodes leads to the topologies shown in \figurename{~\ref{fig:TauTop2}} and \figurename{~\ref{fig:TauTop3}}. Assume the topology is initially as the one in \figurename{~\ref{fig:TauTop2}}:
\begin{itemize}
\item Node $N_1$ unicasts a ${\it rec\_newpkt}({\it data}, N_2)$ to itself, indicating that it wants to send $\it data$ to node $N_2$.
\item
Node $N_1$ initiates a route discovery procedure by broadcasting an
${\it rreq}_{N_1,0}$ message to its neighbors, i.e., nodes $N_3$ and $N_2$. Note
that ${\it rreq}_{a,i}$ refers to an $\it rreq$ message received from node $a$
with the hop count of $i$. Each $rreq$ message has more parameters but here
only these two parameters are of interest and the other parameters are
assumed to be equal for all the $rreq$ messages, i.e., the destination and
source sequence numbers, and the source and destination IP addresses.
\item
Node $N_3$ processes the ${\it rreq}_{N_1,0}$ and since it is not the
destination and has no route to $N_2$ in its routing table,
rebroadcasts the ${\it rreq}_{N_3,1}$ message to its neighbors, nodes $N_1$
and $N_2$, after increasing the hop count. At this point, node $N_2$
has two messages in its queue, ${\it rreq}_{N_1,0}$ and ${\it rreq}_{N_3,1}$.
\item
Node $N_1$ moves out of the communication range of node $N_2$,
resulting the network topology shown in
\figurename{~\ref{fig:TauTop3}}.
\item
Node $N_2$ takes ${\it rreq}_{N_1,0}$ from the head of its queue and updates
its routing table by setting $N_1$ as the next hop in the route
towards $N_1$. As node $N_2$ is the intended destination for the route
discovery message, it unicasts an $\it rrep$ message towards the
originator, $N_1$, indicating that the route has been built and it
can start forwarding the data. Therefore, node $N_2$ attempts to
unicast an $\it rrep$ message to node $N_1$, i.e., its next hop towards the
originator.
\item Since the connection between the nodes $N_1$ and $N_2$ is broken, it fails to receive an ack from $N_1$ and marks the route as \emph{invalid}.
\item Node $N_2$ takes ${\it rreq}_{N_3,1}$ from its queue and since the route state towards $N_1$ is \emph{invalid}, it evaluates the received route to determine whether it is loop free. Updating the routing table with the received route is said to be
``loop free", if the received message cost, e.g., the hop
count is less than or equal to the existing route cost. Since the
hop count of the received message is greater than the existing one,
it does not update the existing route and the message is discarded.
\end{itemize}

Although the route through node $N_3$ to node $N_1$ seems to be valid,
the protocol refuses to employ it to prevent possible loop formation in
the future. %AODV uses timer to rebroadcast $rreq$ messages if no
%$rrep$ was received.

%% file: RelatedWork.tex
\section{Related Work} \label{sec::related}
A large number of studies has been conducted for modeling and
verification of MANET protocols using different approaches to tackle
its specific challenges. These challenges, as discussed in Section
\ref{sec::wrebec}, are modeling the underlying topology, mobility and
local broadcast.

Some works model and analyze the correctness of MANET protocols using existing formal frameworks such as
SPIN~\cite{de2004formal,wibling2004automatized} and
UPPAAL~\cite{fehnker2012automated,mclver2006formal,wibling2005ad}. In a SPIN model, node connectivity is modeled with the help of PROMELA channels, one for each node. Also, mobility is modeled by \textit{case selection} instruction provided by PROMELA, for modeling nondeterminism. In the initialization section, possible links to other neighbors are defined as different \textsf{case}s that all will be checked for a model. Since it does not provide a specific technique to reduce the state space, its state space grows very fast and it is only feasible to check small topologies. 
Therefore, models would be limited to fewer nodes. In UPPAAL, connectivity is modeled through a set of arrays of booleans, while changing topology is modeled by a separate automaton which manipulates the arrays. In
\cite{wibling2004automatized}, a case study was carried out to evaluate two model checkers, SPIN and UPPAAL. Due to state-space explosion, the analysis was limited to some special mobility scenarios (as a part of the specification). However, our reduction technique makes it possible to verify a MANET for all possible topology changes to find an error.

As
explained in~\cite{ene99expressiveness}, from a theoretical point
view, compositionality is not preserved if broadcast is encoded based on
point-to-point communications. Lack of support for
compositional modeling and arbitrary topology changes has motivated
new approaches with a primitive for local broadcast and support of
arbitrary mobility in an algebraic way. These approaches include CBS\#
\cite{nanz2006framework}, CWS \cite{Mezzetti2006331}, CMN
\cite{Merro2009194}, the $\omega$-calculus \cite{Singh2010440},
bA$\pi$ \cite{godskesen2010observables}, CMAN
\cite{godskesen2007calculus,Godskesen09}, RBPT
\cite{ghassemi2008restricted} and the bpsi-calculi
\cite{borgstrom2013broadcast,psi}. Each of these proposed frameworks
overcome the modeling difficulties such as local broadcast and its message delivery guarantee property and mobility in different ways.
%They usually truly overcome one or two of  these challenges while
%the others remain unresolved. 
All these approaches, except
CBS\#, CWS, AWN and bpsi-calculi, suffer from lack of message delivery
guarantee that makes them inappropriate for analyzing properties
such as packet delivery \cite{fehnker2013process}. They model broadcast
through either an enforced  synchronized or lossy communication. When communications are
lossy, a node may not receive a message although it is in the transmission range of the
sender. CBS\# and CWS use enforced
synchronization for broadcast to make sure that all ready nodes
within the transmission range of a sender will receive the message.
Although they guarantee message delivery to the ready receivers, it
is not possible to define meaningful nodes (which can successfully receive messages while they are processing another message) in their syntax which are
always ready (i.e, {\it input-enabled}) \cite{fehnker2013process}. 
The process algebra
AWN is proposed particularly for modeling wireless mesh network
(WMN) routing protocols which uses local broadcast with message
delivery guarantee. It defines its own data structures to model
routing tables and other necessary data types to model the AODV protocol.
In addition, conditional unicast is introduced for modeling the
procedure to act based on the message delivery acknowledgment. In all
these approaches, while a node is busy processing a message, it
fails to receive messages from other nodes. Therefore, either nodes
are defined to be input-enabled at the semantics as in CBS\# and CWS or a process with a
queue that concurrently stores new messages should be specified at the syntax as in AWN and bpsi-calculus.
Almost all these languages model mobility of nodes in their semantics through arbitrary changes
of the topology with the exception that it is modeled through different generations of assertions on connectivity information in \cite{borgstrom2013broadcast}.  
%Therefore, bringing the
%queue at specification level prevents from using reduction
%techniques like counting abstraction. 
In wRebeca,
communications are asynchronous and received messages are stored in
queues implicitly at the semantic level (without the need to make nodes explicitly input-enabled). %This also
%not only abstract the
%delay of network (not in order delivery), but also 
%enables us to easily
%exploit the counter abstraction technique. 
Furthermore, the atomic execution of message handlers, which avoids unnecessary interleaving of the node behaviors, together with topology abstraction through $\tau$-elimination technique, where the topology changes are a source of state-space explosion in the process calculi approaches, make our framework applicable to the verification of real-world yet complex protocols such as AODV. %Our semantics (after
%$\tau$-transition removal) is a CLTS similar to RBPT with the
%difference that states are counted together, resulting a more
%compact CLTS.
We remark that unnecessary interleaving of behaviors can be handled in bpsi-calculus by means of priorities.

%As we
%restrict our models to the ones with a fixed and known number of
%nodes, they boil down to the finite state problems. 
There are different approaches
\cite{AlParam,Param,DAbdulla} with the aim to analyze
networks with an infinite number of nodes, where nodes execute an
instance of a network process. A network configuration is
represented as a graph in which each individual node represents a state of
the process. The behavior of a process is modeled by an automaton.
The network configuration transforms due to either the process
evolution at a network node or the topology reconfiguration.
Verification of safety properties, reaching to an undesirable
configuration starting from an initial configuration, is
parameterized due to any possible number of nodes and connectivity
among them. It is proved that the problem of parameterized safety
properties, the so-called \textit{control states reachability problem},
is undecidable. However, that problem turns out to be decidable for
the class of bounded path graphs \cite{Param,AlParam}. Decidability
of the problem was also considered when configurations evolve due to
discrete/continuous clocks at processes \cite{DAbdulla}. Furthermore, an
inductive approach based on reduction to prove compositional
invariants for the dynamic process networks was presented in \cite{Inductive}. This approach reduces the calculation of a compositional invariant to a smallest representative network through setting up a collection of local symmetry relations between nodes, specifically defined for each problem. The computed non-dynamic compositional invariant on the representative network is generalized for the entire dynamic network family when the non-dynamic invariant is preserved by any reaction to a dynamic change in the network. 
%This techniques define how to prove a property for all network and processes. It discusses parametric verification in presence of adversy which chagnes the links. To this aim, compostional invariant verification is defined in terms of local invariants which should be preserved by transitions, and interferece of two neighbours, and addition/removal of links, its consequnce interference. IT use localized neighbour symmetries to minimize calculation of compositional invarinat to a set of representative processes
%Their symmetry reduction exploits localized neighborhood symmetries
%defined as a relation among the local states of nodes which is preserved by transitions, and interference of two neighbors. 
%
Then, they proved loop freedom of AODVv2-04
for an arbitrary number of nodes in \cite{LoopNamj} through an inductive and compositional proof: It provides an inductive invariant and proves that it is held initially and also preserved by every action, either a protocol action or a change in the network, similar to the approach of \cite{Dist}. They have reported two loop-formation scenarios due to inappropriate setting of timing constants and accepting any valid route when the current route is broken without any further evaluation (to ensure loop formation). Another approach is based on graph transformation systems, where network configurations are hypergraphs and transitions are specified by graph rewriting rules, modeling the dynamic behavior of a protocol. Safety properties are symbolically specified by graph patterns, a generalized form of hypergraphs with negative conditions, through a \textit{symbolic
backward reachability analysis} which is not
guaranteed to terminate due to the undecidability of the problem
\cite{SaksenaWJ08}. To this aim, an over-approximation of the set of configurations preceding a bad configuration are computed by using a fixed point analysis, and then check that this set contains no
initial configuration. While these approaches are scalable to prove a
property for MANETs with a potentially unbounded number of nodes with an exhaustive effort, our approach is valuable to easily examine
confirmation and diagnostics of suspected errors in the early phase
of protocol development for a limited number of nodes. In other
words, our efficient model checking tool can be used as an initial
step before involving to generalize a property for an arbitrary
sized network.

%% file: Conclusion.tex
\section{Conclusion and Future Work}
In this paper we extended the syntax and semantics of bRebeca, the
actor-based modeling language for broadcasting environment, to
support wireless communication in a dynamic environment. We
addressed the key features of wireless ad hoc networks, namely reliable
local broadcast, conditional unicast, and last but not least
mobility. The reliable asynchronous local broadcast/unicast
communication, and implicit support of message storages 
% and effect
%of network delays on message delivery, 
make our framework suitable
to analyze MANETs with respect to
different mobility scenarios. A modeler only focuses on how to
decompose a protocol into a set of communicating actors to cover
functionalities of the protocol under investigation.

To overcome the state-space explosion, we leveraged the counter
abstraction technique to analyze ad hoc networks with static
topologies. Our reduction technique performs well on protocols with no
specific state variable that distinguishes each rebec, and topologies
with many topologically equivalent nodes. We demonstrated the effectiveness
of our approach on the flooding protocol in different network
settings. However, mobility ruins the soundness of our counting
abstraction. To this end, we eliminated $\tau$-transitions while
topology information was removed from the global states to considerably
reduce the size of the state space. We integrated the proposed reduction
techniques into a tool customizable in verifying wRebeca models for
different message storage policies and the topology dynamism. Invariants
can be checked during the state-space generation while the resulting output
can be fed into the existing model checking tools such as mCRL2 and CLTS
model checker.

%Furthermore, the reduced semantic model
%is supported by a model checker tool to verify the
%topology-dependant behavior of MANETs.

We presented a complete and accurate model of the core
functionalities of a recent version of AODVv2 protocol (version 11).
We abstracted optional features and timing aspects to make our model
manageable. We verified the loop freedom property in AODVv2-11 and found
a scenario in which the property violated. The scenario was confirmed by the AODV group. Loop freedom has already
been proved on
various versions of AODV: AODVv1-02 \cite{BhargavanOG02}, AODV-rfc3561 \cite{fehnker2013process,Dist}, and AODVv2-04 \cite{LoopNamj,SaksenaWJ08}, respectively. Among these only \cite{LoopNamj} considers the timed behavior of the AODV. The new version differs in the following
aspects which distinguish our attempt: in this version multiple next
hops are maintained for each destination and consequently the
process to update the routing table is completely different;
Different statuses are considered for a route in the table of a node regarding the
neighbor status of its next hop; Sequence numbers for invalid destinations
in intermediate nodes are not increased anymore (like
\cite{LoopNamj,SaksenaWJ08}, in contrast to others). Although, these approaches
focus on providing a general proof for the property, our model
checking-based approach detects the error and the scenario that
leads to it. Our approach, can be adopted to resolve
conceptual/design errors in an iterative way in the early phase of
protocol development. The positive result of verifications constitutes a predicate about the protocol for a
limited number of nodes. The combination of model checking and theorem proving techniques allows to prove a predicate about a MANET protocol for any number of nodes.

We plan to integrate our state-space generator tool into the
verification environment \cite{afra} to take advantage of its model
checker. Furthermore, we aim to run more cast studies to extend
application of our framework. To analyze real-time and probabilistic
behaviors of wireless network protocols, wRebeca can be extended in
the same way of \cite{KhamespanahFTTS,varshosaz2012modeling}. To
this aim, there is a need to examine the soundness of our reduction techniques when probability and time are introduced.

\section*{Acknowledgements}
We would like to
thank the anonymous reviewers for their constructive comments on the earlier version of the paper, the AODV group for their supports, Mohammad Reza Mousavi for his discussion on the paper, Wan Fokkink and Bas Luttik for their helpful comments on the paper.